\documentclass[12pt, nofootinbib]{article}

\usepackage{amssymb}
\usepackage{amsmath,bm}
\usepackage{amssymb}
\usepackage{graphicx}
\usepackage{amsfonts}         
\usepackage{fancybox}

\usepackage{enumitem}

\usepackage{slashed}

\usepackage[font=small,labelfont=bf]{caption}
\DeclareCaptionFont{tiny}{\tiny}
\usepackage{yfonts} 
\usepackage{epstopdf}
\usepackage[usenames,dvipsnames]{xcolor}
\usepackage[pagebackref,  
pdftex, bookmarks={false}, pdfauthor={John McGreevy}, pdftitle={Yay, physics!}]{hyperref}
\hypersetup{colorlinks=true, 
linkcolor=BrickRed, 
citecolor=Violet, 
filecolor=OliveGreen, 
urlcolor=RoyalBlue, 
filebordercolor={.8 .8 1}, 
urlbordercolor={.8 .8 0}}
\usepackage{soul}
\setstcolor{Red}

\usepackage{wrapfig}

\definecolor{darkgreen}{rgb}{0,0.4,0}
\definecolor{darkred}{rgb}{0.4,0,0}
\definecolor{darkblue}{rgb}{0,0,0.4}
\definecolor{lightblue}{rgb}{.6,.6,0.9}

\newcommand{\cor}{\color{red}}

\newcommand{\cob}{\color{black}}

\definecolor{uglybrown}{rgb}{0.8,  0.7,  0.5}

\definecolor{palatinatepurple}{rgb}{0.41, 0.16, 0.38}
\definecolor{celebrationcolor}{rgb}{0.75,  0.0,  0.9}

 \usepackage{mdframed}

\usepackage{framed}
\definecolor{shadecolor}{rgb}{0.90,0.90,0.90}

\usepackage{wasysym}

\def\parfig#1#2{
\parbox{#1\textwidth}
{\includegraphics[width=#1\textwidth]{#2}}
}

\numberwithin{equation}{section}

\renewcommand{\theequation}{\arabic{section}.\arabic{equation}}

\def\nd{{ \vphantom{\dagger}}}

\input epsf

\newlength{\extraspace}
\newlength{\extraspaces}
\def\be{\begin{equation}}
\def\ee{\end{equation}}

\newcommand{\bea}{\begin{eqnarray}}
\newcommand{\eea}{\end{eqnarray}}

\def\tr{{\rm tr}}

\def\bra#1{\left\langle#1\right|}
\def\ket#1{\left|#1\right\rangle}

\def\CE{{\cal E}}

\def\CH{{\cal H}}
\def\CI{{\cal I}}

\def\CL{{\cal L}}

\def\CN{{\cal N}}
\def\CO{{\cal O}}

\def\CR{{\cal R}}
\def\CS{{\cal S}}

\def\II{\relax{I\kern-.10em I}}

\def\IB{\relax{\rm I\kern-.18em B}}

\def\ID{\relax{\rm I\kern-.18em D}}
\def\IE{\relax{\rm I\kern-.18em E}}
\def\IF{\relax{\rm I\kern-.18em F}}
\def\IG{\relax\hbox{$\inbar\kern-.3em{\rm G}$}}
\def\IGa{\relax\hbox{${\rm I}\kern-.18em\Gamma$}}
\def\IH{\relax{\rm I\kern-.18em H}}
\def\II{\relax{\rm I\kern-.18em I}}
\def\IK{\relax{\rm I\kern-.18em K}}

\def\inbar{\,\vrule height1.5ex width.4pt depth0pt}

\def\IR{\mathbb{R}}

\def\lp10{\ell_p^{10}}
\def\lp11{\ell_p^{11}}
\def\R11{R_{11}}

\def\frac#1#2{{#1 \over #2}}

\newdimen\tableauside\tableauside=1.0ex
\newdimen\tableaurule\tableaurule=0.4pt
\newdimen\tableaustep
\def\phantomhrule#1{\hbox{\vbox to0pt{\hrule height\tableaurule width#1\vss}}}
\def\phantomvrule#1{\vbox{\hbox to0pt{\vrule width\tableaurule height#1\hss}}}
\def\sqr{\vbox{%
  \phantomhrule\tableaustep
  \hbox{\phantomvrule\tableaustep\kern\tableaustep\phantomvrule\tableaustep}%
  \hbox{\vbox{\phantomhrule\tableauside}\kern-\tableaurule}}}
\def\squares#1{\hbox{\count0=#1\noindent\loop\sqr
  \advance\count0 by-1 \ifnum\count0>0\repeat}}
\def\tableau#1{\vcenter{\offinterlineskip
  \tableaustep=\tableauside\advance\tableaustep by-\tableaurule
  \kern\normallineskip\hbox
    {\kern\normallineskip\vbox
      {\gettableau#1 0 }%
     \kern\normallineskip\kern\tableaurule}%
  \kern\normallineskip\kern\tableaurule}}
\def\gettableau#1 {\ifnum#1=0\let\next=\null\else
  \squares{#1}\let\next=\gettableau\fi\next}

 \def\eqnn#1{\xdef #1{(\secsym\the\meqno)}\writedef{#1\leftbracket#1}%
 \global\advance\meqno by1\wrlabeL#1}
 \def\eqna#1{\xdef #1##1{\hbox{$(\secsym\the\meqno##1)$}}
 \writedef{#1\numbersign1\leftbracket#1{\numbersign1}}%
 \global\advance\meqno by1\wrlabeL{#1$\{\}$}}
 \def\eqn#1#2{\xdef #1{(\secsym\the\meqno)}\writedef{#1\leftbracket#1}%
 \global\advance\meqno by1$$#2\eqno#1\eqlabeL#1$$}

\global\newcount\itemno \global\itemno=0

\def\itemaut#1{\global\advance\itemno by1\noindent\item{\the\itemno.}#1}

\def\({\left(}
\def\){\right)}

\def\rrho{{\boldsymbol\rho}}

\def\lsim{\mathrel{\mathstrut\smash{\ooalign{\raise2.5pt\hbox{$<$}\cr\lower2.5pt\hbox{$\sim$}}}}}
\def\gsim{\mathrel{\mathstrut\smash{\ooalign{\raise2.5pt\hbox{$>$}\cr\lower2.5pt\hbox{$\sim$}}}}}

\def\overleftrightarrow#1{\vbox{\ialign{##\crcr
     $\leftrightarrow$\crcr\noalign{\kern-0pt\nointerlineskip}
     $\hfil\displaystyle{#1}\hfil$\crcr}}}
     
     \def\overleftarrow#1{\vbox{\ialign{##\crcr
     $\leftarrow$\crcr\noalign{\kern-0pt\nointerlineskip}
     $\hfil\displaystyle{#1}\hfil$\crcr}}}

\def\eg{{\it e.g.}}
\def\ie{{\it i.e.}}

\def\gSU{\textsf{SU}}

\usepackage[yyyymmdd,hhmmss]{datetime}
\newif{\ifeq}           
\eqtrue                 
\newcounter{lecturecounter}

\renewcommand{\title}[1]{\vbox{\center\LARGE{#1}}\vspace{5mm}}
\renewcommand{\author}[1]{\vbox{\center#1}\vspace{5mm}}
\newcommand{\address}[1]{\vbox{\center\em#1}}

\renewcommand{\date}[1]{\vbox{\center#1}}

\usepackage{multirow}
\usepackage{geometry}
\usepackage{pdflscape}

\newcommand{\beq}{\begin{equation}}
\newcommand{\eeq}{\end{equation}}

\newtheorem{DEF}{Definition}

\newcommand{\MI}{\mathcal{I}}

\begin{document}

\begin{titlepage}

\title{{\Huge Mixed $s$-sourcery:}\\
Building many-body states using bubbles of Nothing
}

\author{Brian Swingle${}^a$ and John McGreevy${}^b$}

\address{${}^a$ Department of Physics, Stanford University, Palo Alto, CA 94305, USA}

\address{${}^b$ Department of Physics, University of California at San Diego, La Jolla, CA 92093, USA}

\begin{abstract}
We recently introduced the idea of $s$-sourcery \cite{sourcery1}, a general formalism
for building many-body quantum ground states using renormalization-group-inspired quantum circuits.
Here we define a generalized notion of $s$-sourcery that applies to mixed states, and study its properties and applicability. For our examples we focus on thermal states of local Hamiltonians. We prove a number of theorems establishing the prevalence of mixed $s$-source fixed points, giving results for free fermion models, conformal field theories, holographic models, and topological phases. Thermal double states (also called thermofield double states) and the machinery of approximate conditional independence are used heavily in the constructions. For a large class of models we provide an information theoretic argument for the existence of a local Hamiltonian whose ground state is the thermal double state, and in some cases we construct such a Hamiltonian.

\end{abstract}

\vfill

\today

\end{titlepage}

\vfill\eject
\setcounter{tocdepth}{1}    

\tableofcontents

\vfill\eject

\section{Introduction and definitions}

Many interesting states of matter cannot
be built by acting on a product state with a local unitary circuit of small depth.
In \cite{sourcery1}, we introduced a quantitative refinement
to this obstruction, via a program which we call $s$-sourcery.
The idea is:
rather than building the state all at once, to build the system hierarchically, one scale at a time,
with local unitaries.
The label $s$ specifies the number of
copies of the groundstate of the system required
to double the system size with local unitaries.
These unitaries act on $s$ copies of the system of linear size $L$,
and the requisite number of decoupled degrees of freedom (ancillas),
initialized in product states, to produce a single copy of the system of linear size $2L$.

A more highly-entangled groundstate requires larger $s$.
(It is conceivable that other resources in the groundstate besides entanglement may
also require larger $s$.)
In \cite{sourcery1}, we used this point of view to prove the area law
for entanglement entropy of subregions for a large class of states matter, namely those
in $d$ spatial dimensions with $ s < 2^{d-1}$.

A unitary map which doubles the system size in this way is called a renormalization group (RG) circuit.
A practical benefit to finding an RG circuit
is that it can be used to construct a MERA
(multiscale entanglement renormalization ansatz) network \cite{mera},
an efficiently-contractible tensor network representation of the groundstate in question.

In this paper, we introduce the idea of mixed-state $s$-sourcery. In particular, the goal is to extend the
$s$-sourcery construction of pure states of Ref.~\cite{sourcery1} to mixed states.
We will focus primarily on thermal states of local Hamiltonians. A key message emerging from these results is that ground state techniques permit us to address thermal states as well. Hence the ground state problem is rather more general than it might at first appear.

{\bf Motivation:} One motivation for the extension of s-sourcery to mixed states comes from hopes of
improving our understanding of transport of charge and energy and of non-equilibrium steady states
in strongly-correlated quantum many-body systems (\eg~\cite{RevModPhys.82.1743, Prosen:2015xha}).
We are particularly interested in cases where a quasiparticle description is not applicable. In such cases, a tensor network representation of the full non-equilibrium steady state might enable efficient calculation of currents and other physical properties. In a forthcoming companion paper, some of the results in this work will be applied to that problem to establish the existence of efficiently contractible tensor networks for a broad class of non-equilibrium steady states \cite{transport-to-appear}.

A skeptical reader might ask:
is not the entanglement structure a delicate groundstate phenomenon which
will be washed out by finite temperature or by coupling to a noisy environment?
Why do we need quantum mechanics in such a situation?

Indeed, in standard examples, one does not need to account for
long-range entanglement\footnote{See, however, \S\ref{sec:long-range-entanglement-at-finite-T}.}.  However,
the form of the short-range entanglement is not altered by
a low-enough temperature, and it is crucial for the
physics. For example, hydrodynamics (a classical description) may correctly describe the long-wavelength and low-energy physics of a strongly interacting conformal field theory at finite temperature, but the transport coefficients entering the hydrodynamic equations contain physics at and above the thermal scale where quantum fluctuations remain important.

To be more precise, consider the density matrix for a locally thermal state of some quantum fluid,
$$ \rho \simeq Z^{-1} e^{ - \sum_x \beta_x \( H_x - J_x v_x \) } ,$$
with $H_x$ the Hamiltonian density, $\beta_x^{-1}$ a local temperature, $J_x$ a current density, and $v_x$ a local fluid velocity. It is assumed that $\beta_x$ and $v_x$ vary slowly on the scale of the thermal correlation length $\xi$. In the worst case, computing averages in such a state is just as hard as
a computing averages in a quantum many-body groundstate. (It may be interesting to formulate a rigorous statement along these lines.)
In this language, the purpose of this line of work is to show that states like
$\rho$ are not worst-case examples, but rather that they are well-approximated by
a certain form, with no long-range entanglement,
which correctly captures the short-range entanglement.
This form is essentially a tensor network.
Analogous to the way hydrodynamics separates low- and high-energy physical processes, this representation {\it factors out the quantum mechanics}:
entanglement and quantum fluctuations are confined to clusters of size $\xi^d$.

Previous progress towards efficient representations of
thermal states of local quantum many body systems
include
\cite{2014arXiv1409.3435K} (in the case of Hamiltonians made from commuting terms)
and \cite{2015PhRvB..91d5138M}. In this context, one of our main results is that a wide variety of thermal states have purifications which can be prepared with a finite depth quantum circuit. This implies that thermal expectation values can be efficiently computed, which is stronger than saying that such states have efficient representions (quasi-polynomial in system size number of parameters).

Another motivation comes from the study of holographic duality, specifically recent proposals relating the circuit complexity of the boundary quantum state to the geometry of the bulk black hole \cite{Stanford:2014jda, Brown:2015lvg, Brown:2015bva}. A major open question there concerns the complexity of the thermal double state. This is a state $\ket{T}_{12}$ on two copies of the system such such $\text{tr}_2(\ket{T} \bra{T}_{12} )  = e^{- H_1/T}/Z$, \ie~$\ket{T}$ is a purification of the thermal density matrix. Such a state is supposed to be dual to the so-called maximal analytic extension of the AdS-Schwarzchild black hole geometry \cite{Maldacena:2001kr}.
It describes an entangled state of two conformal field theories. Our results below provide the first rigorous bounds on the complexity of such states.

A final, more abstract, motivation for this work is the general question: can ground state techniques be usefully applied to thermal states? We argue that the answer is emphatically yes, and in particular, that purifications of thermal states can often be cast as ground states of local Hamiltonians\footnote{Previous work
in this direction is \cite{2013arXiv1308.0756F}.}. To facilitate our analysis, we also make heavy use of a relatively new technology in quantum many-body physics, the machinery of approximate conditional independence describing approximate quantum Markov states (see \eg~\cite{2012PhRvB..86x5116K, 2013PhRvB..87o5120K, fr1, fr2}). Such states have a special structure of correlations that generalizes the classical notion of a Markov chain \cite{qmarkov}, \eg~three systems $A$, $B$, and $C$ such that $A$ and $C$ are independent given $B$. In such cases, correlations between $A$ and $C$ are in essence mediated by $B$, and the global state of $ABC$ takes a particularly simple form. We show that wide variety of thermal states of local Hamiltonians are of this type.

{\bf Plan:} In the remainder of this section we introduce several compelling notions
of what it means for
a density matrix to be an
$s$-source fixed point and
discuss their properties, as well as the relations between these definitions.
In \S\ref{sec:wormhole-argument-here} we describe a strategy
for constructing $s=0$ RG circuits using arrays of holes in space.
\S\ref{sec:free-fermions} proves that a thermal state of a local free fermion hamiltonian
is an $s=0$ fixed point.
\S\ref{sec:CFT} uses our knowledge of the entanglement entropy of subregions
in a thermal state of a conformal field theory (CFT) to
constrain its mixed-$s$-sourcery index (\ie~$s$).
For the special case of CFTs with a classical gravity dual,
\S\ref{sec:holography} shows that
$s=0$ is built into the Ryu-Takayanagi formula \cite{2006PhRvL..96r1602R}.
\S\ref{sec:RGdecomposition} incorporates
our understanding of $s>0$ groundstates
to improve the locality properties of the quantum circuits resulting from our construction. \S\ref{sec:gauge-theory}
analyzes topological gauge theories
in various dimensions as $s$-source
fixed points, at $T=0$ and $T>0$. This analysis leads us to conjecture that mixed-$s>0$ is a condition
for finite-temperature quantum memory.
\S\ref{sec:open-questions} discusses new frontiers opened by
this line of work and several applications of our results. The appendices collect some useful background material.

\subsection{Definition of mixed s sourcery}

In this section we present three definitions of mixed state s sourcery in order of increasing generality. The most general definition turns out to be the easiest to work with, at least as far as the computations in this paper are concerned, but all three seem to be useful for studying many-body mixed states. Also, all three forms have the property that correlations may be efficiently computed.  We will discuss the relationships between the various definitions afterward.
First, we define the basic notion of a quasi-local quantum channel.

A quasi-local quantum channel is a completely positive trace preserving map $\mathcal{N}$ from states of system $A$ to states of system $A'$ with certain locality properties. It must be of the form
\beq
\mathcal{N}(\rho_A) = \tr_{(A')^c} \left(U^\dagger_{AE} \rho_A \sigma_E U_{AE} \right),
\eeq
where an environment $E$ is introduced, the trace is over the complement $(A')^c$ of $A'$ in $AE$ (we allow $A'$ to be bigger than $A$ so only a part of $E$ is traced out), $\sigma_E$ is a fixed local product state, and $U_{AE}$ is a quasi-local unitary. In short, we may tensor in extra degrees of freedom in a fixed local product state, act with a quasi-local unitary, and then trace some of the extra degrees of freedom back out. We often refer to such maps as local channels. It will sometimes be necessary to specify the range $r$ of the quasi-local unitary (length scale beyond which the terms in $U_{AE}$ decay faster than any power) in which case we refer to an $r$-local channel.

\begin{DEF}[Mixed $s$ sourcery: Noisy]
A sequence of states $\{ \rho_L\}$ form a \textbf{noisy} $s$ source fixed point if
\beq
\rho_{2L} = V\(\underbrace{\rho_L \otimes ... \otimes \rho_L}_{s \,\text{times}} \otimes \sigma\)V^\dagger
\eeq
where $\sigma$ is a product state (the noise) and $V$ is a quasi-local unitary.
\end{DEF}

With the noisy definition the relationship between different scales is required to be unitary, but we are allowed to inject extra noise in the form of the state $\sigma$. However, no extra degrees of freedom are allowed to be traced out so we may think of this as a case where an environment is introduced which cannot be traced out. This is the most restrictive definition of mixed-state $s$ sourcery.
Related definitions have been suggested in \cite{Swingle:2009bg, Swingle:2012wq} and in \cite{2014arXiv1412.0732E}.

\begin{DEF}[Mixed $s$ sourcery: Open]
A sequence of states $\{ \rho_L\}$ form an \textbf{open} $s$ source fixed point if
\beq
\rho_{2L} = \CN\left(\underbrace{\rho_L\otimes  ... \otimes \rho_L}_{\text{$s$ times}}\right)
\eeq
where $\CN$ is a quasi-local quantum channel.
\end{DEF}

With the open definition the relationship between different scales is no longer unitary but is allowed to be a more general quantum operation. The difference between the open and noisy definitions is simply that in the open definition we are allowed to trace out part of the environment. Every noisy $s$ source fixed point is also an open $s$ source fixed point; for example, we can generate the noise $\sigma$ using a local channel and then apply the unitary transformation and the composition of these two transformations is a quasi-local channel.

\begin{DEF}[Mixed $s$ sourcery: Purified]
A sequence of states $\{ \rho_L\}$ form a \textbf{purified} $s$ source fixed point if there exists a sequence of purifications $\{\ket{\sqrt{\rho_L}}_{12}\}$ with $\tr_2( \ket{\sqrt{\rho_L}} \bra{\sqrt{\rho_L}}_{12})= \rho_L$ and
\beq
\ket{\sqrt{\rho_{2L}}} = \tilde{V}
\(  \underbrace{\ket{\sqrt{\rho_L}}\otimes ... \otimes \ket{\sqrt{\rho_L}}}_{\text{$s$ times}} \otimes \ket{0...0} \)
\eeq
where $\ket{0...0}$ is a product state of the appropriate size and $\tilde{V}$ is a quasi-local unitary on $A^sE$.
\end{DEF}

A purified $s$ source fixed point is the most general form we consider in this paper. Any open $s$ source fixed point is also a purified $s$ source fixed point since we may simply hold onto the environment which purifies the action of the channel at each step.
The difference is that in the purified case, we do not reuse
an assumed-to-be-forgetful environment at each step;
the cost is a geometric buildup of the local dimension,
of order the full Hilbert space dimension.\footnote{For example, every channel from $n$ to $n$ qubits can be obtained by a unitary acting on $n+2n$ qubits (the $2n$ comes from the environment). Since the number of qubits at step $\ell$ of the growth process is $2^{d\ell}$, the size of the environment at every growth step is proportional $2^{d\ell}$. The geometric sum $\sum_\ell 2^{d\ell}$ counts the total number of environment qubits used and is of order $2^{d \ell_{\max}}$ where $N = 2^{d \ell_{\max}}$ is the final number of qubits.}

When the sequence of states $\{\rho_L\}$ are thermal states of a local Hamiltonian, we will often refer to the purification $\ket{\sqrt{\rho_L}}_{12}$ as a thermal double state, and we denote it as $\ket{T}$. (In the field theory literature such a state is also called a thermofield double state, but we prefer the simpler moniker.) The thermal double state is not unique: if $\ket{T}_{12}$ is any thermal double state, then $I_1 \otimes U_2 \ket{T}_{12}$ is also a thermal double state (with respect to system $1$).

Some comments follow:
\begin{itemize}
\item As in \cite{sourcery1}, we are using the phrase ``fixed point" loosely to refer to the entire phase.
\item When making quantitative error estimates, we should specify a maximum range for the local channels and/or local unitaries in question.
\item More generally, the value of $s$ may be scale-dependent.
Depending on the range of the unitaries, some states, \eg, a $d=1$ free fermion metal at low $T$, will be $s=1$ until the scale of $\xi=v_F/T$ ($v_F$ is the Fermi velocity) is reached at which point it becomes $s=0$.
\item The states $\{\rho_L\}$ could be thermal states of the same Hamiltonian on different system sizes or it could be useful to let the Hamiltonian flow.
\item Finally, the above definitions include and generalize the ground state definition given in
\cite{sourcery1}.
\end{itemize}

We note that Hastings has proposed a definition of topological order at finite temperature \cite{2011PhRvL.107u0501H}: a state is not topologically ordered
when it can be obtained from a classical state (a state diagonal in a local product basis) by acting with a local channel. However, because a classical state
can be long-range correlated, Hastings' notion of triviality is in fact
stronger than the $s=0$ condition: $s=0$ states are topologically trivial by Hastings' definition, but
the reverse is not necessarily true.

\subsection{Wormhole arrays and bubble-of-Nothing arrays}
\label{sec:wormhole-argument-here}

In this section we explain the origin of our subtitle and discuss in broad terms our approach. First we discuss the scope of the work and then transition into a discussion of the key idea of bubble-of-Nothing arrays.

Although we focus on non-zero temperature states for most of the paper, mixed state $s$ sourcery can be useful even for zero temperature pure states; in this sense the word ``mixed" is a bit of a misnomer, we are just using the distinction between a local channel and a local unitary even when the output is pure. The following example shows that, at least in the special case of ground states which are pure, there is a distinction between between noisy and open fixed points.

Symmetry-protected states are zero temperature quantum phases that are non-trivial only given a certain symmetry; here non-trivial means that the system must pass through a phase transition to reach a trivial product state. Symmetry-protected states are defined to be short-range entangled in the sense of having no topological entanglement entropy, no anyonic excitations or excitations with fractional quantum numbers, and no topological ground state degeneracy. However, not all symmetry-protected states can be produced from product states using a short depth circuit. For example, integer quantum Hall states (which are in fact not protected by any symmetry but are simply protected) cannot be produced from product states using a short depth circuit. Hence they cannot be noisy $s=0$ fixed points (the ``noise" would have to be pure in this case). Nevertheless, all symmetry-protected states, including those which are protected in the absence of any symmetry, are open $s=0$ fixed points.

The proof is simple: All such states $\ket{\psi}$ are invertible \cite{sourcery1}, which means there is an inverse state $\ket{\psi^{-1}}$ such that we can produce $|\psi\rangle |\psi^{-1}\rangle$ from a product state with a quasi-local unitary. The channel then consists of an environment which duplicates the original system, the above quasi-local unitary which produces $\ket{\psi}\ket{\psi^{-1}}$, and a tracing out of the environment. Since the system and environment are not entangled after the quasi-local unitary acts, the output is a pure state of the system.

In a little more detail, the notion of invertible states in \cite{sourcery1} is based on the following idea: the state of a gapped system may be deformed into a product state by first deforming the state to introduce an array of small holes into the system and then expanding the holes until they consume the entire system. If both steps can be done without closing the gap, then the state of the system is equivalent to a product state\footnote{In the case of symmetry-protected states, the process should also respect the given symmetry at every step,
so that the resulting circuit commutes with the symmetry.}. This was called the wormhole array argument in \cite{sourcery1} but
from the point of view of the state $\ket{\psi}$, it could be called the bubble-of-Nothing array argument. It is similar to an independent unpublished construction due to Kitaev \cite{Kitaev-2013}.

\begin{figure}[htbp]
\includegraphics[width=\textwidth]{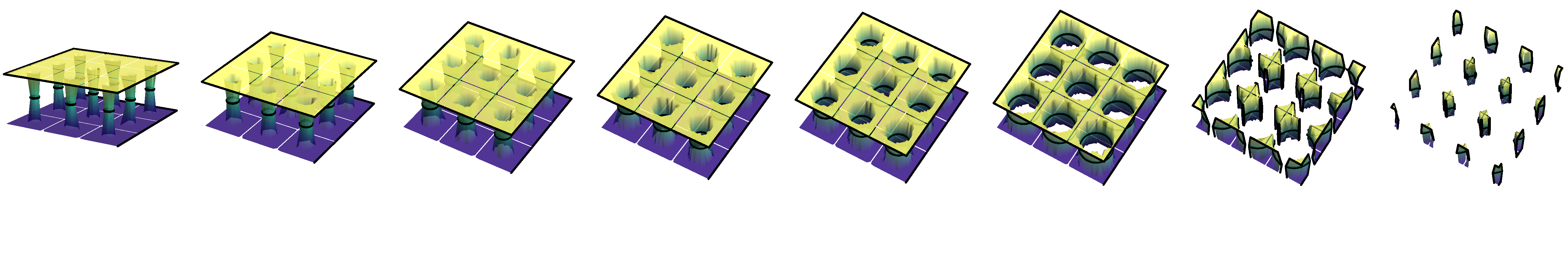}
\caption{\label{fig:wormhole-array}  A 2d state and its inverse state, connected by an array
of wormholes.  The wormholes expand until the system is decomposed into an array of isolated
molecules.}
\end{figure}

Given a gapped topological liquid, there are two possible obstructions to deforming the system
in the manner we require, one associated with long-range entanglement and one associated with gapless edge or boundary states.
For symmetry-protected states, long-range entanglement is ruled out by the assumption of the vanishing of the topological entropy, the absence of anyons, etc. (we analyze this obstruction in
more detail in \S\ref{sec:gauge-theory}).
But even in the absence of long-range entanglement, the system can possess edge states which make the system with holes gapless. This prevents the holes from being adiabatically expanded. However, as argued in \cite{sourcery1}, every short-range entangled state has an inverse state such that the combined system has no protected edge states. The coupling of the system and its inverse can be depicted as a wormhole connecting the system and its inverse, so the array of holes is understood as an array of wormholes in this case, as depicted in Fig.~\ref{fig:wormhole-array}.

The wormhole/bubble-of-Nothing array argument can be described in precise terms as a construction or deconstruction of the state using a cellular decomposition of space. For example, starting from the full state in $d$ dimensions we first remove the $d$-cells to leave a $(d-1)$-skeleton. Then we remove the $(d-1)$-cells to leave a $(d-2)$-skeleton, and so forth. At the end of the procedure we are left with decoupled $0$-cells in a product state.  The unitary time evolution which reverses
this procedure gives the desired hierarchical construction of the state
from product states ($s=0$).

The preceding analysis was for gapped ground states of local Hamiltonians, but very similar reasoning applies to short-range correlated thermal states. In our discussion of CFTs at finite temperature
\S\ref{sec:CFT} we will give a detailed discussion of the cellular construction of thermal states. We also discuss in more detail
in \S\ref{sec:gauge-theory}
the precise manner in which long range entanglement represents a topological obstruction
to this kind of hierarchical reconstruction from product states.
As we will see, a virtue of this bubble-of-Nothing array argument is that it exposes physics
content of the topological entanglement entropy (TEE) terms in dimensions higher than two, \ie~gives them an operational meaning.

\subsection{Properties}

Here we record some properties of mixed $s$ source fixed points.

\subsubsection{Equivalence of open and purified for $s=0$}

We have seen that the noisy and open definitions need not be be equivalent even at $s=0$, \ie~there are $s=0$ open fixed points that are not also $s=0$ noisy fixed points. However, the purified and open definitions do agree at $s=0$.

\subsubsection{Mutual information growth}

While a local channel can produce volume law entropy, it cannot produce much correlation and entanglement as measured by the mutual information $I(A,B) \equiv S(A) + S(B) - S(AB)$. Given a subregion $A_R$ of size $R$ in $d$ spatial dimensions it is true that
\beq
I(A_{2R},A_{2R}^c) \leq s I(A_R,A_R^c) + k R^{d-1}.
\eeq
The constant $k$ depends on the shape and is proportional to the logarithm of the local dimension (having in mind that the system is defined on a lattice or graph with a fixed local Hilbert space).

The mutual information bound is proven using the Stinespring dilation of the local channel which performs the mapping. Fix a region of interest (and use the channel variables introduced above). The initial mutual information of the enlarged system is $s$ times the mutual information of the chosen region (because initially we tensor in a product state $\sigma_E$). The final mutual information of the enlarged system is upper bounded by the initial mutual information plus an area law's worth of mutual information. This is because the mutual information of $U^\dagger_{AE} \rho_A \sigma_E U_{AE}$ is bounded by the mutual information of $\rho_A$ plus an area law contribution due to $U_{AE}$ ($\sigma_E$ being a product state) \cite{2013PhRvL.111q0501V}. Finally, since mutual information decreases when regions are discarded, $I(A,BC)\geq I(A,B)$, it follows that the mutual information of the final state after the partial trace over $E$ is still bounded as claimed.

\subsubsection{Induced map on local operators}

Second, a local channel maps local operators to local operators. Any channel has a (non-unique) Kraus representation as
\beq
\mathcal{N}(\rho) = \sum_\alpha M_\alpha \rho M_\alpha^\dagger
\eeq
where
\beq
\sum_\alpha M_\alpha^\dagger M_\alpha = 1.
\eeq
By switching to the ``Heisenberg picture" for channels, we may write
\beq
\tr\left(O \mathcal{N}(\rho)\right) = \tr\left(\mathcal{N}^\dagger(O) \rho \right) \equiv \sum_\alpha \tr\left(M_\alpha^\dagger O M_\alpha \rho\right).
\eeq
If the channel is local then the Kraus operators may be chosen to be local\footnote{Proof:  The channel is local means that it has a purification
with a local unitary $U$.  But then Kraus operators can be taken to be
$ M_\alpha = \bra{\alpha} U \ket{0} $
where $\ket{0}$ is a reference state of the environment,
and $\ket{\alpha}$ is a large enough set of basis states of the environment.
Taking matrix elements on the environment preserves the range of the operator on the system,
as long as the basis is local.}, so the new operator $\mathcal{N}^\dagger(O)$ is local if the old operator $O$ is local. Hence expectation values may still be computed efficiently.

{\bf A toy example.} Consider a linear array of $L$ qubits. Suppose each qubit experiences a dephasing channel of the form
\beq\label{eq:dephasing}
\mathcal{D}(\rho) = (1-p)\rho + p Z \rho Z.
\eeq
This channel has Kraus operators $M_1 = \sqrt{1-p}$ and $M_2 = \sqrt{p} Z$. The channel acting on the entire qubit array is $\mathcal{D}^{\otimes L}$ and has $2^L$ Kraus operators including
\beq
\sqrt{1-p}^k \sqrt{p}^{L-k} 1 \otimes ... \otimes 1 \otimes Z \otimes ... \otimes Z
\eeq
for all $k$ and all possible permutations of the operators.

The conjugate channel $\left(\mathcal{D}^{\otimes L}\right)^\dagger$, when acting on a local operator $O$, will have most of the Kraus operators pair up as $M_\alpha^\dagger M_\alpha$ and sum to $1$ away from $O$. Near $O$ a few remaining Kraus operators don't precisely cancel and we obtain a new local operator as the output of the conjugate channel acting on $O$.

\section{Free fermions at finite temperature are $s=0$}
\label{sec:free-fermions}

In this section we prove the following theorem: given any quadratic local fermion Hamiltonian of the form $H = \sum_{x,y} c_x^\dagger h_{x,y} c_y^\nd$ ($x$ and $y$ are position labels, $h_{x,y}$ decays as $|x-y| \rightarrow \infty$), any thermal state of $H$ with $T>0$ is an $s=0$ purified fixed point. The range of the required local channel is given by the correlation length of the finite temperature state. The proof uses the machinery of thermal double states (also called thermofield double states), so we first introduce the necessary background.

Given a local Hamiltonian $H$ the thermal state at temperature $T$ is
\beq
\rho(T) = \frac{e^{-H/T}}{Z}
\eeq
where $Z(T) =\tr\left(e^{-H/T}\right)$ is the partition function. This mixed state of the system may be derived from a pure state by introducing a second copy of the system. The thermal double state,
\beq
|T\rangle_{12} = \sum_E \sqrt{\frac{e^{-E/T}}{Z}} |E\rangle_1 |E\rangle_2,
\eeq
has the property that
\beq
\tr_2\left(|T\rangle \langle T |\right) = \rho_1 = \rho(T).
\eeq
There are in fact many thermal double states; one infinite family is obtained by taking $\ket{T}_{12}$ and acting with any unitary of the form $I_1\otimes U_2$. It is sometimes convenient to partially fix this freedom by demanding that $\ket{T}_{12}$ be a $+1$ eigenstate of a swap operator which exchanges the two systems.

The key physical idea is this: because $|T\rangle$ has short-range correlations, one might suspect that it could be construed as the ground state of a gapped Hamiltonian.
We now show that this is the case for free fermion thermal states; a hamiltonian whose ground state is $\ket{T}$ will be called a thermal double Hamiltonian. The proof that all finite temperature free fermion states are purified $s=0$ fixed points then follows by constructing a family of thermal double Hamiltonian which interpolate between temperature $T$ and infinite temperature (where the state is ultra-local).

For convenience we describe the construction in one spatial dimension and for spinless fermions. None of these simplifications are essential and the theorem is completely general. The only assumptions are $T>0$ and a local quadratic Hamiltonian.

Consider a one-dimensional translation invariant chain of fermions with Hamiltonian
\beq
H = \sum_k \epsilon_k c_k^\dagger c_k^\nd.
\eeq
The momentum space operators $c_k$ are related to position space operators via
\beq
c_k = \sum_{x} \frac{e^{-ikx}}{\sqrt{L}} c_x
\eeq
with $L$ the number of sites. Whether the spectrum of $H$ is gapped or gapless above the ground state, the corresponding thermal state has decaying correlations. Introduce a second copy of the system with fermion operators $\tilde{c}_k$.

To get the main idea, focus for a moment on a single mode $c$ with energy $\epsilon$. This mode should be occupied with probability $f(\epsilon) = (e^{\epsilon/T}+1)^{-1}$, so we may write the relevant thermal double state as
\beq
|T\rangle = \sqrt{1-f} |n = 0\rangle |\tilde{n} = 1\rangle + \sqrt{f} |n=1\rangle |\tilde{n}=0\rangle
\eeq
where $n = c^\dagger c$ and $\tilde{n} = \tilde{c}^\dagger \tilde{c}$. Note that we have implemented a convenient particle-hole transformation so that each state in the superposition has the same charge (same eigenvalue of $n+\tilde{n}$). It is convenient to defined rotated modes (not to be confused with the dimension of space),
\beq
d = \sqrt{f} c + \sqrt{1-f} \tilde{c}
\eeq
and
\beq
\tilde{d} = -\sqrt{1-f} c + \sqrt{f} \tilde{c},
\eeq
so that the thermal double state is simply $|T\rangle = d^\dagger |\text{vac}\rangle$.

The thermal double state is trivially the ground state of a Hamiltonian of the form
\beq
h_T = - d^\dagger d + \tilde{d}^\dagger \tilde{d}.
\eeq
Indeed, the ground state of this Hamiltonian simply has $d$ occupied and $\tilde{d}$ empty. Notice also that the gap is independent of $f$, but of course locality has no meaning in this single mode example.

Now we generalize this construction to the multi-mode system described by $H$. Let $f_k$ be the average occupation of level $\epsilon_k$ (Fermi function) and define
\beq
d_k = \sqrt{f_k} c_k + \sqrt{1-f_k} \tilde{c}_k
\eeq
and
\beq
\tilde{d}_k = -\sqrt{1-f_k} c_k + \sqrt{f_k} \tilde{c}_k.
\eeq
The thermal double Hamiltonian is taken to be
\beq
H_T = \sum_k \left(- d^\dagger_k d_k^\nd + \tilde{d}^\dagger_k \tilde{d}_k^\nd \right),
\eeq
so that its ground state,
\beq
|T\rangle = \prod_k d_k^\dagger |\text{vac}\rangle,
\eeq
is the thermal double state for $\rho(T)$.

As in the single mode case, the gap of $H_T$ is constant independent of $T$ and system size $L$. However, it is not obvious that $H_T$ is a local Hamiltonian. We now show that it is. Expand $H_T$ in terms of the original modes to obtain
\beq
H_T = \sum_k \left((1-2 f_k) c^\dagger_k c_k^\nd - (1-2 f_k) \tilde{c}^\dagger \tilde{c}^\nd - 2 \sqrt{f_k(1-f_k)} (c_k^\dagger \tilde{c}_k^\nd + \tilde{c}^\dagger_k c_k^\nd)\right).
\eeq
Expand each $c_k$ and $\tilde{c}_k$ in terms of position space operators to construct the real space representation of $H_T$,
\beq
H_T = \sum_{x,y} \left( J_1(x-y) c_x^\dagger c_y^\nd - J_1(x-y) \tilde{c}_x^\dagger \tilde{c}_y^\nd - J_2(x-y)(c_x^\dagger \tilde{c}_y^\nd + \tilde{c}^\dagger_y c_x^\nd) \right),
\eeq
where
\beq
J_1(x-y) = \frac{1}{L}\sum_k e^{ik(x-y)} (1-2f_k)
\eeq
and
\beq
J_2(x-y) = \frac{1}{L} \sum_k e^{ik(x-y)} 2 \sqrt{f_k (1-f_k)}.
\eeq

Now $J_1$ is essentially just the correlation function of $c_x^\dagger c_y^\nd$ so it is exponentially decaying by assumption. $J_2$ is a little more complex, but it is also the Fourier transform of a smooth function (provided $T>0$), so it too will decay faster than any power of $|x-y|^{-1}$. In fact, both functions will decay exponentially with some correlation length set by some combination of the energy gap (if it exists) and the temperature. For example, a finite temperature metallic state would have $\xi(T) = v_F/T$ where $v_F = \partial_k \epsilon_k |_{k_F}$. Hence $H_T$ is a quasi-local Hamiltonian with range $\xi$ (the correlation length).

At infinite temperature all the $f_k = 1/2$, so $H_T$ takes a particularly simple form
\beq
H_{\infty} = - \sum_x ( c_x^\dagger \tilde{c}_x^\nd + \tilde{c}^\dagger_x c_x^\nd )
\eeq
which is ultra-local. The ground state of this Hamiltonian is manifestly a product state in position space and can be obtained from a product state of the $c$ and $\tilde{c}$ variables by the action of an ultra-local unitary which simply adds one particle in the mode $(c_x+\tilde{c}_x)/\sqrt{2}$ for each $x$.

Since the gap of $H_T$ is independent of $T$ and the range of $H_T$ is bounded for any non-zero $T$, it follows that there is a quasi-local unitary transformation mapping the ground state of $H_T$ to the ground state of $H_\infty$. To construct this unitary we use quasi-adiabatic continuation \cite{2005PhRvB..72d5141H, 2004PhRvB..69j4431H,2010arXiv1008.5137H}. Introduce a family of Hamiltonians $H(\eta) = H_{T/\eta}$ with the property that $H(1) = H_T$ and $H(0)= H_\infty$. $H(\eta)$ has a uniform gap and range bounded by $\xi(T)$ (assuming $\xi(T)$ is a decreasing function of $T$), so the quasi-adiabatic generator
\beq
-i K(\eta) = \int_{-\infty}^\infty dt F(t) e^{i H(\eta) t} \partial_\eta H(\eta) e^{-i H(\eta) t}
\eeq
($F$ is the usual filter function) is local by Lieb-Robinson.

Thus the ground state of $H_T$ is related to a product state by quasi-local unitary as claimed. The thermal state is obtained by tracing out the auxiliary $\tilde{c}$ system. This establishes the theorem. Furthermore, the proof is constructive: the quasi-local generator $K(\eta)$ is explicitly computable from $H(\eta)$ once $F$ is fixed.

{\bf A comment on locality.} In our explicit free fermion construction the range of the thermal double Hamiltonian directly corresponded to the range of correlations in the thermal state. This enabled the gap of the thermal double Hamiltonian to be constant as a function of $T$. We might ask an alternative question: is there a thermal double Hamiltonian where fixed range interactions but with a gap that depends on $T$ and on the gap of the original Hamiltonian? Such a construction would accommodate long-range correlations not by increasing the range of a fixed gap Hamiltonian but by decreasing the gap of a fixed range Hamiltonian.

It is not at all clear that this can be accomplished in general, although it is amusing to note that the $T=0$ state, where correlations might actually be long-ranged, is by assumption the ground state of a local Hamiltonian.

In the case of free fermions one can make partial progress as follows. By expanding the function $f(\epsilon)$ in powers of $\epsilon/T$ one can correctly capture the low energy part of the thermal state (meaning the occupation numbers of states near $\epsilon=0$ is correctly reproduced), but the high energy states are not correctly captured. In particular, states very far below and above the Fermi level are much more softly occupied than in the true thermal state. Depending on the question, this approximation may or may not be valuable. In any event, using $1-2f \approx \frac{\epsilon}{2T}$ and $\sqrt{f(1-f)} \approx \frac{1}{2}$ we see that the effective single particle Hamiltonian
\beq
h_k = \left(
        \begin{array}{cc}
          \frac{\epsilon_k}{2T} & \frac{1}{2} \\
          \frac{1}{2} & -\frac{\epsilon_k}{2T} \\
        \end{array}
      \right)
\eeq
is local with a velocity set by $T$. Alternatively, by multiplying through by $T$ (which does not change the ground state), we obtain a Hamiltonian with a gap proportional to $T$. The approximation may be improved by keeping more terms in the expansion of $f$ at the cost of increased non-locality.

{\bf Case of free bosons.} We explicitly studied the case of free fermions, but free bosons yield the same result. Consider a single bosonic mode $b$ obeying $[b,b^\dagger]=1$ with Hamiltonian $H= \omega b^\dagger b$. The thermal state is
\beq
\rho(T) = \sum_{n=0}^\infty \frac{e^{-\beta \omega n}}{Z} \ket{n} \bra{n}
\eeq
with $Z = (1- e^{-\beta \omega})^{-1}$. Introducing a second mode $\tilde{b}$, the thermal double state is
\beq
\ket{T} = \frac{1}{Z} \sum_{n=0}^\infty e^{-\beta \omega n /2 } \ket{n}_b \ket{n}_{\tilde{b}}.
\eeq

This thermal double state can be interpreted as a two-mode squeezed state,
\beq
\ket{T} = e^{- r b \tilde{b} + r b^\dagger \tilde{b}^\dagger } \ket{0} \ket{0},
\eeq
where $\tanh r = e^{- \beta \omega /2}$. One can further verify that the two-mode squeezed state is the ground state of a Hamiltonian quadratic in $b$ and $\tilde{b}$ (this is because the squeezing unitary implements a linear transformation among $b$, $b^\dagger$, $\tilde{b}$, and $\tilde{b}^\dagger$), so the remainder of the fermion analysis can be immediately applied.

\section{CFTs at finite temperature are $s=0$}
\label{sec:CFT}

In this section we establish the existence of an approximate local channel for general thermal states of field theories. It is inspired by the wormhole array argument
described in \S\ref{sec:wormhole-argument-here}
and by Petz's reconstruction map for states saturating strong sub-additivity \cite{qmarkov}.
Here by CFT, we intend also gapped CFTs, \ie~topological field theories.
Essentially we aim to reconstruct the CFT thermal state from local data using a sequence of local quantum channels.
We prove the following theorem: the thermal state of any CFT which is smoothly connected to $T=\infty$
 is an open $s=0$ fixed point, where the local channel has a range set by the thermal length.

The argument is valid for known CFTs in $d=1,2,3$ but can fail in higher dimensions due to a topological obstruction to reconstruction which persists at finite temperature. We discuss these topological obstructions in more detail later. For now, note that most quantum critical points of physical interest have $d<4$ and the arguments given in this section apply.

The key input needed for the construction is the form of the entanglement entropy for various regions.  Given a CFT in $d+1$ spacetime dimensions, the construction succeeds if the entropy of any region $A$ of linear size $\ell$ has the form
\beq
\label{eq:entropyform}
S(A) = c_1 |A| + \int_{\partial A} \(c_2 + \sum_{i>2} c_i f_i( K, R) \) + \CO(\ell^d e^{-\ell/\xi}).
\eeq
The first term is a volume term, the second term is a local integral over $\partial A$ of polynomials of local curvatures (extrinsic $K$ and intrinsic $R$), and the final term is an exponentially small correction (provided $\ell \gg \xi$). If this form is obeyed, then appropriate linear combinations of entropies will cancel the volume and boundary terms leaving only exponentially small corrections.

The form \eqref{eq:entropyform} holds if the thermal state in question
is adiabatically connected to infinite temperature.
That's because at high temperature,
the state may be constructed by a path integral where the thermal
circle is much smaller than other length scales, including the size of the region;
such a thin path integral produces an $S(A)$ of the form \eqref{eq:entropyform},
and a phase transition is required to change this form.

Zero temperature gapped states also typically have the form \eqref{eq:entropyform} of entanglement with $c_1 = 0$, but with additional sub-leading topological terms. Such terms, if present, interfere with the cancellation of various entropies and can obstruct the reconstruction from local data.
When these terms survive to finite temperature they may obstruct the reconstruction of the finite temperature state from local data.
Constraints on the associated crossover functions were studied in
\cite{2013Swingle-Senthil}.

We note that if there is a pure state
with the same entanglement structure
of not-too-large subregions as the thermal state
(as would follow if highly excited pure states of the system generically locally thermalize),
then this forbids the terms which are odd under the orientation reversal of the boundary,
such as odd powers of the extrinsic curvature.

In the remainder of this section we assume \eqref{eq:entropyform} and
use it to construct the $s$-sourcery map.
In the next section we show that \eqref{eq:entropyform} holds for
holographic CFTs.
We return to its possible violation in section \S\ref{sec:gauge-theory}.

Briefly, the key idea is that the form \eqref{eq:entropyform}
implies that the components of a cellular
decomposition of space (as in the bubble-of-Nothing construction)
form an approximate quantum Markov chain (see \S\ref{subsec:qmarkov} for a brief review).
The newly-developed understanding
of the consequences of such approximate conditional independence
\cite{fr1, fr2} then implies the existence of the required $s=0$ channel.

\subsection{CFT$_{1+1}$}

As an example consider a conformal field theory in $1+1$ dimensions. The entropy of an interval of length $\ell$ is (see \eg~\cite{Calabrese:2009qy})
\beq
S(\ell) = \frac{c}{3} \log\left(\frac{\sinh(\pi T \ell)}{\pi T a}\right)
\eeq
where $a$ is a UV cutoff. One easily checks that $S(\ell)$ has the claimed form with $c_1 = \pi c T/3$ and $c_2 = \frac{c}{3} \log\left(\frac{\xi}{a}\right)$ (there is no curvature term) when $\ell \gg 1/T = 2\pi \xi$.

\begin{figure}
\centering
\includegraphics[width=.8\textwidth]{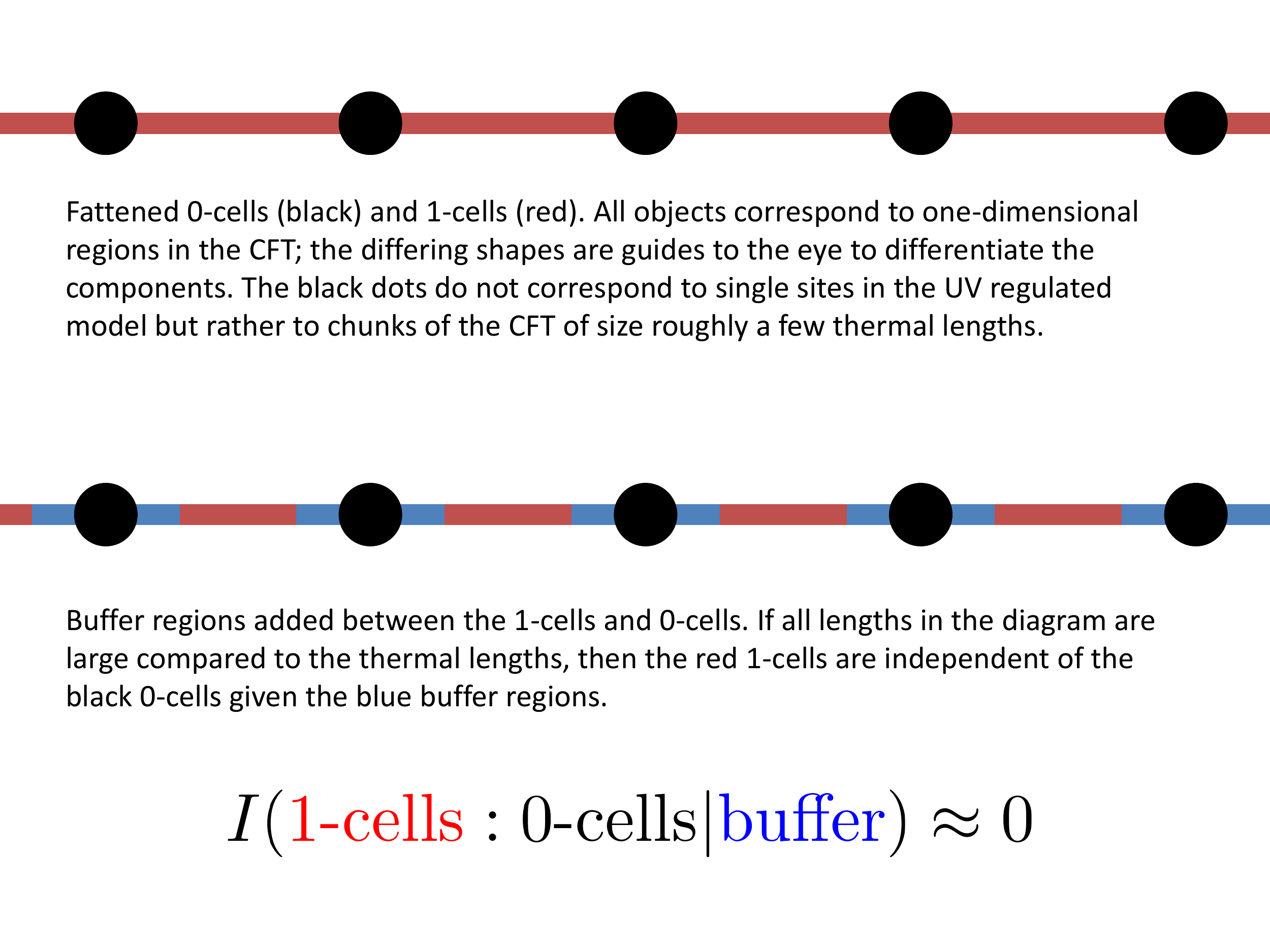}
\caption{\label{markov}
Geometry of approximate conditional independence calculation in $d=1$.}
\end{figure}

Now consider a cellular decomposition of space. We have thickened $0$-cells of size $\ell_0$ and $1$-cells of length $\ell_1$. We also allow a buffer region of size $\ell_{b}$. All lengths are taken to be larger than the correlation length and will be specified in more detail shortly. The decomposition is shown in Fig.~\ref{markov} where the $0$-cells are in black, the $1$-cells are in red, and the buffer region is blue.

The idea of the conditional independence construction is to show that the $1$-cells are independent of the $0$-cells given the buffer regions. If this is true then we can stitch space together using local channels because the state is an approximate quantum Markov chain.

The quantity we need to evaluate is $\CI = I(1\text{-cells}:0\text{-cells}|\text{buffer})$ where $I(A:C|B)$ is the conditional mutual information of $A$ and $C$ given $B$ which takes the form $I(A:C|B) = S(AB)+S(BC)-S(B)-S(ABC)$.

Since all lengths are assumed larger than $\xi(T) = \frac{1}{2\pi T}$, the asymptotic form of $S(A)$ given in \eqref{eq:entropyform} applies.\footnote{The calculation $S(\text{many intervals})$ needed to obtain $\CI$ may be performed using the operator product expansion for the correlator of twist fields on the torus. The geometry is such that the dominant contribution comes from each pair of twist fields (associated to a single interval) fusing to the identity.} First, observe that the volume terms, $c_1 |A|$, cancel in $\CI$ since we have $|AB| +|BC|-|B|-|ABC| =0$. Similarly, the area terms, $c_2 |\partial A|$, also cancel since the number of positive boundaries cancels the number of negative boundaries. This means that the only contribution which doesn't cancel is the exponentially suppressed terms, $\mathcal{O}(\ell^{d} e^{-\ell/\xi})$. Hence we find that
\beq
\MI = \mathcal{O}(N_{\text{cells}} e^{-\ell/\xi})
\eeq
where $\ell = \min\{\ell_0,\ell_1,\ell_b\}$ and $N_{\text{cells}}$ is the number of $1$-cells (proportional to system size $L$).

Demanding that $\CI < \epsilon$ requires taking $\ell = \xi \log\left(\frac{N_{\text{cells}}}{\epsilon}\right)$ which is only modestly larger than the correlation length even for very large system sizes. \cite{fr1} then implies that there exists a recovery channel which reconstructs $\rho_{1\text{-cells},0\text{-cells},\text{buffer}}$ from $\rho_{0\text{-cells},\text{buffer}}$ (which is approximately a product state) using a local channel acting just on the buffer region and the $1$-cells. \cite{fr2} shows that this channel is independent of the state of the $0$-cells. The $0$-cells are themselves obtainable from a local channel. Hence the entire thermal state of the CFT can be constructed from the composition of two channels, one which instantiates the $0$-cells and one which builds in the $1$-cells. This means that the thermal state of the CFT is approximately $s=0$.

What is the actual channel? Both \cite{fr1} and \cite{fr2} are non-constructive, but it is known that for pure states the transpose channel (see \S\ref{subsec:reverse}) is within a square root of being optimal \cite{prettygoodchannel}. Running the same argument for a thermal double state then allows the use of the transpose channel as the approximate reconstruction channel with a worse error of $\sqrt{\epsilon}$ instead of $\epsilon$ (but this is a very modest worsening since the range depends only on $\log \epsilon$). Building on more recent results, we give an explicit formula for a channel with the desired properties in the discussion at the end of the paper. We also note that because the mutual information approximately vanishes between the relevant regions, \eg~between different parts of the buffer, the channels may all be taken to be local.\footnote{This mutual information condition is non-trivial. Consider the state $\ket{0...0}\bra{0...0} + \ket{1...1}\bra{1...1}$ and let $A$, $B$, and $C$ be the cellular regions discussed above and shown in Fig.~\ref{markov}. This state has the property that the entropy of any region is $1$, so $I(A:C|B) = 0$. However, for any subregions $b$ and $b'$ of the buffer $I(b:b') \neq 0$, so while there is a channel to reconstruct $ABC$ from $AB$ and $BC$, this channel does not factorize over the different buffer regions in Fig.~\ref{markov}. Indeed, there are just two Kraus operators which measure $\{\ket{0},\ket{1}\}$ and reattach the pieces as appropriate.}

We conjecture that this approximate local reconstruction map can be completed to a quasi-local channel which exactly reproduces the thermal state. This would imply that all thermal states of CFTs are $s=0$ (as expected). In the case of free fermion CFTs we can prove this statement (previous section).

\subsection{CFT$_{d+1}$}

The above argument generalizes immediately to higher dimensions provided the higher dimensional analog of \eqref{eq:entropyform} is obeyed. The general procedure is to consider a fattened cellular decomposition of space and assemble the state one step at a time. For typical CFTs the conditional informations still approximately vanish.\footnote{This can again be seen from an operator product expansion point of view except that now we must deal with extended twist fields \cite{Swingle:2010jz, Hung:2014npa, Hung:2011nu, Bueno:2015qya}. The dominant fusion channel is still to the identity, that is, the twist field of each region prefers to annihilate with itself rather than pairing with another twist field.} However, we must be cautious in higher dimensions because of the possibility of corrections to \eqref{eq:entropyform} coming from topological terms which persist to finite temperature. For example, the $2$-form toric code in $d=4$ at low temperatures has such obstructions \cite{Dennis:2001nw}, as we discuss in detail below. This obstruction means that the $2$-form toric code remains $s=1$ at finite temperature.

A more innocuous new wrinkle in $d>1$ is the possibility of regions whose
boundaries have sharp corners.  Such corners produce
additional universal singular terms \cite{Bueno:2015rda, Bueno:2015qya}
which depend on the geometry of the corner.
Like the (non-universal) area terms, these corner terms cancel pairwise
in the conditional mutual information
of the subcells. To be more precise, if the system is gapless, then at zero temperature these corner terms can involve logarithms of various subsystem sizes. Such logarithms would only be expected to cancel up to an order one constant; this is not sufficient for our construction. However, if at finite temperature there is a decay of correlations set by a thermal length, then the corner terms should involve logarithms of the thermal length (instead of the subsystem size) up to exponentially small corrections. In this case, our claimed cancellation is valid for sufficiently large regions.

\begin{figure}[h]
\begin{center}
\includegraphics[width=.326\textwidth]{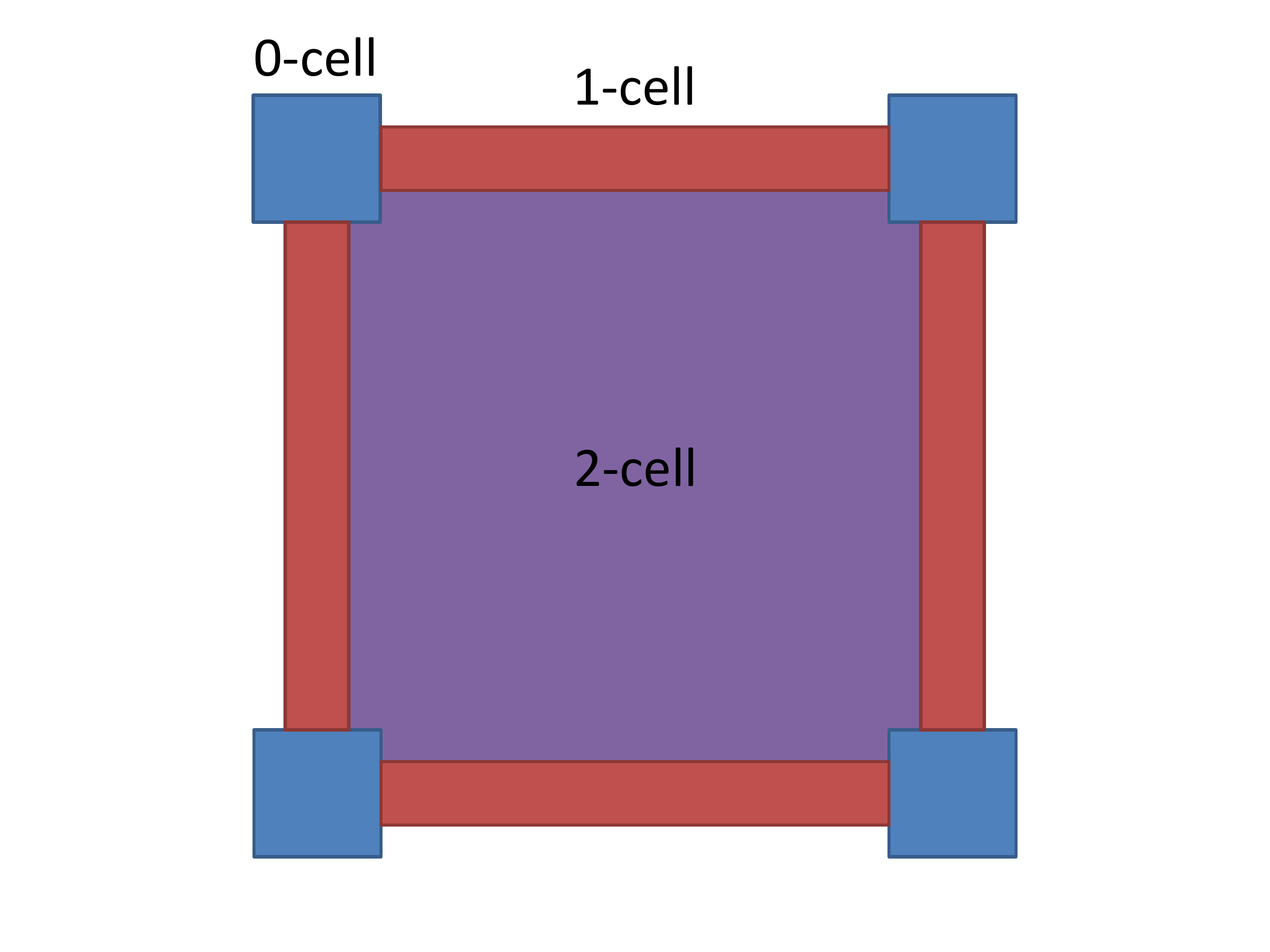}
\includegraphics[width=.326\textwidth]{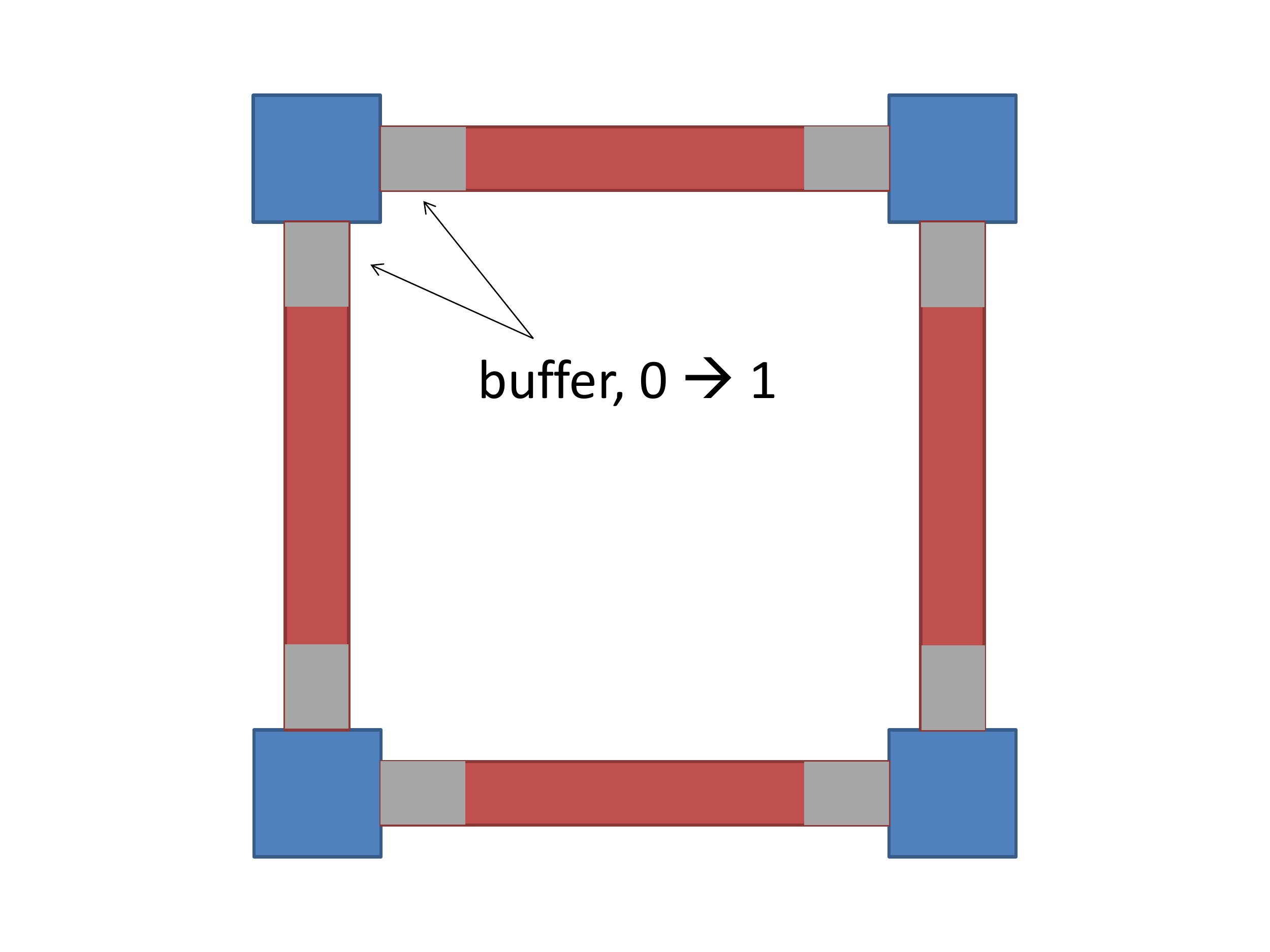}
\includegraphics[width=.326\textwidth]{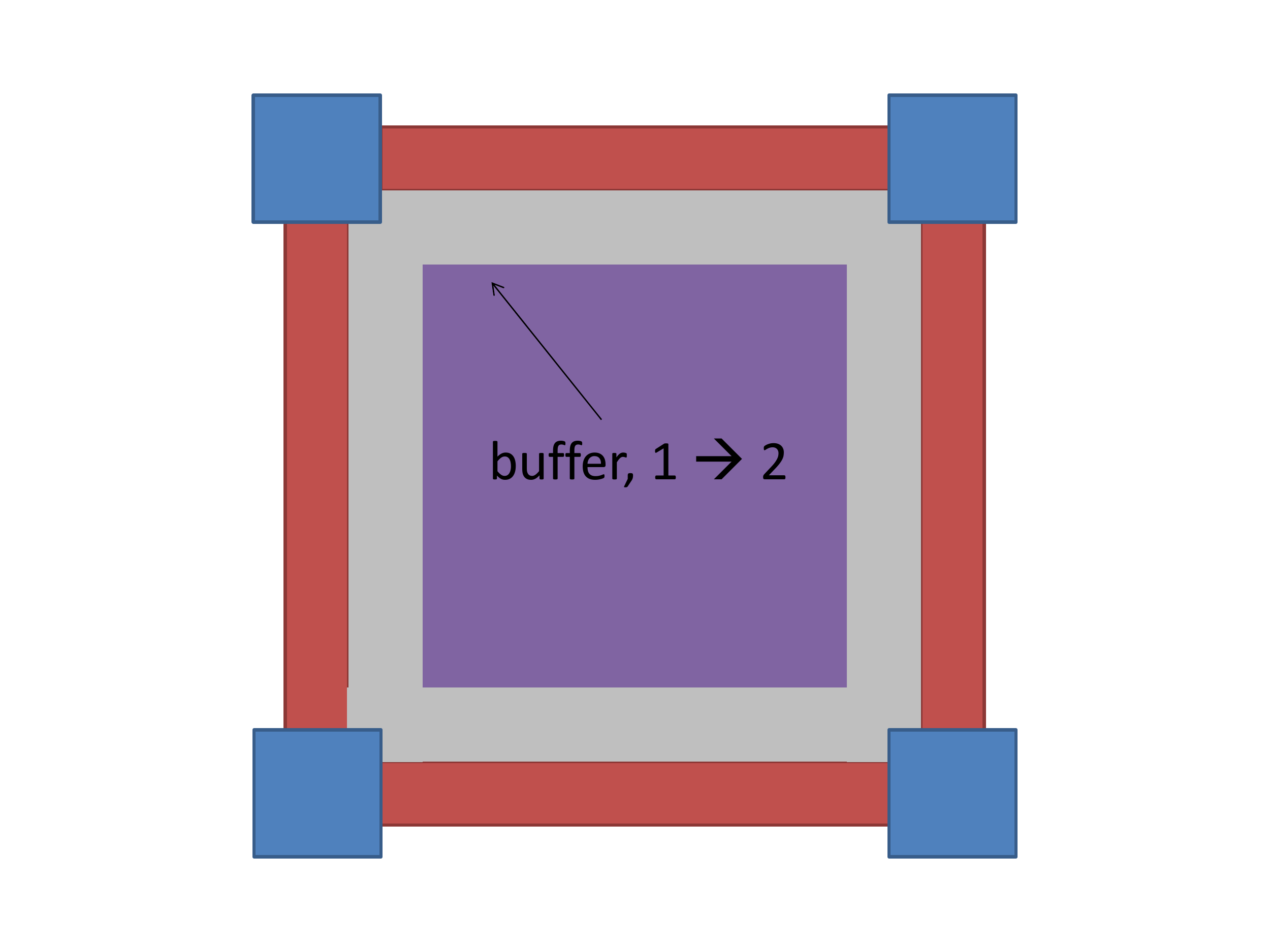}
\caption{A region of the cellular decomposition in two dimensions.  The middle figure shows
the `buffer' regions between the 0-cells and the 1-cells.
The right figure shows the buffers between the 1-cells and 2-cells.
\label{fig:2d-cells}
}
\end{center}
\vskip-.2in
\end{figure}

We illustrate the cellular construction in $d=2$; the general construction follows from it in a straightforward way. The geometry is shown in Fig.~\ref{fig:2d-cells}. We begin with $0$-cells which being in a product state can be produced using a local channel, call it $\CN_{\emptyset \rightarrow 0}$, so that $\rho_{\text{$0$-cells}} = \CN_{\emptyset \rightarrow 0}(\cdot)$.
(Note that $\CN_{\emptyset \rightarrow 0}$ is a map from the empty set.)
Next we verify that $\CI_{0\rightarrow 1} = I(\text{$1$-cells}:\text{$0$-cells}|\text{buffer}) \approx 0$ using \eqref{eq:entropyform}. Approximate conditional independent implies that there exists a channel which approximately reconstructs the state $\rho_{\text{$1$-cells}\cup \text{$0$-cells}}$ from $\rho_{\text{$0$-cells}}$ (the buffer is present as well, but we always absorb the buffer into the $1$-cells for the purposes of the latter parts of the construction.) Call the channel $\CN_{0\rightarrow 1}$ so that $\rho_{\text{$1$-cells}\cup \text{$0$-cells}} = \CN_{0\rightarrow 1}(\rho_{\text{$0$-cells}})$. Finally we verify that $I(\text{$2$-cells}:\text{$0$-cells}\cup \text{$1$-cells}|\text{buffer}) \approx 0$ again using \eqref{eq:entropyform}. This gives a final channel $\CN_{1\rightarrow 2}$ which satisfies $\rho_{\text{$2$-cells} \cup \text{$1$-cells}\cup \text{$0$-cells}} = \CN_{1\rightarrow 2}(\rho_{\text{$1$-cells}\cup \text{$0$-cells}})$. The channels $\CN_{\emptyset \rightarrow 0}$, $\CN_{0 \rightarrow 1}$, $\CN_{1\rightarrow 2}$ are local, so the composite channel is also local but with a larger range. Thus we have
\beq
\rho(T) = \rho_{\text{$2$-cells} \cup \text{$1$-cells}\cup \text{$0$-cells}} \approx \CN_{1\rightarrow 2}(\CN_{0\rightarrow 1}(\CN_{\emptyset\rightarrow 0}(\cdot))),
\eeq
so thermal states of CFTs are $s=0$ with range set by the thermal length.

We again conjecture that the local channel which approximately produces the thermal state can be completed to a quasi-local channel that exactly produces the thermal state. This conjecture is true for free fermion CFTs in any dimension. An argument analogous to the free fermion construction can be constructed for finite temperature free boson states as well.

One consequence of our results is that, for any CFT obeying our entropy assumption, there exists a local Hamiltonian whose ground state is approximately the thermal double state. The Hamiltonian is local with a range set by the thermal length (or correlation length). The above general information theoretic arguments do not permit us to further constrain the form of such a thermal double Hamiltonian, although in some special cases, \eg~free fermions, we could say considerable more about the complexity of the local terms. Some additional examples and comments about thermal double Hamiltonians are given in the discussion. Finally, note that a positive answer to our conjecture about the existence of an exact quasi-local channel would imply the existence of a quasi-local Hamiltonian whose exact ground state is the thermal double state.

\begin{figure}[h]
\begin{center}
\includegraphics[width=.5\textwidth]{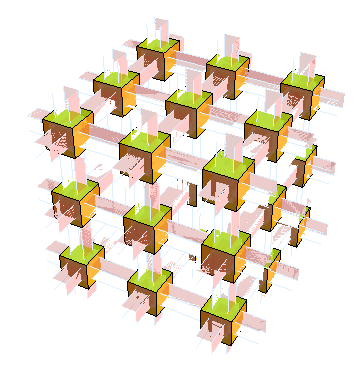}
\caption{The cellular decomposition in three dimensions: the 0-skeleton is
shown as large cubes; the 1-skeleton adds in the pink prisms.
The 2- and 3-cells are not shown.}
\end{center}
\end{figure}

\section{Holographic models at finite $T$ are $s=0$}
\label{sec:holography}

In this section we give a demonstration that holographic models are $s=0$ when the state of the system is such that geometry has an IR wall. This includes black holes, hydrodynamic states in the context of the fluid/gravity correspondence, and gapped states, \eg~the AdS soliton. The approach is again to use approximate conditional independence. The main technical point is that the Ryu-Takayanagi formula \cite{2006PhRvL..96r1602R} enables easy computation of entropies at leading order in large $N$ (gauge theory gauge group is, \eg~$\gSU(N)$). However, we must be a little cautious since an order one term in $I(A:C|B)$ can still obstruct reconstruction. Thus we must also make an argument about the sub-leading terms in the entanglement entropy; the argument uses the Faulkner-Lewkowycz-Maldacena correction term \cite{Faulkner:2013ana}.

The key result from the holographic computations of entropy is that \eqref{eq:entropyform} is again obeyed up to $\CO(1/N)$ terms. Thus in the large $N$ limit the state obeys approximation conditional independence and the cellular decomposition used above provides a framework to construct such a holographic state.

We now explicitly show this for black hole states including an analysis of the sub-leading correction terms. Results for the much more general setting of the fluid/gravity correspondence will be presented in \cite{Hubeny-Swingle}.

{\bf Sketch of holographic duality.} For readers unfamiliar with holographic duality
\cite{1999IJTP...38.1113M,1998PhLB..428..105G, 1998AdTMP...2..253W}
we very briefly sketch the needed ideas. A less compact and more useful discussion can be found in a number of reviews \cite{Aharony:1999ti, Maldacena:2003nj, Horowitz:2006ct, Hartnoll:2009sz,
2009arXiv0909.3553H, McGreevy:2009xe}. First we review the gravitational basics and then briefly discuss the duality and the key dictionary entries translating between bulk and boundary.

For simplicity we consider pure Einstein gravity with negative cosmological constant in $D+1$ dimensions. $D=d+1$ is the spacetime dimension of the boundary field theory. The dynamical field in the bulk is the metric $g_{ab}$; from the metric we can construct various geometrical objects including the Riemann curvature tensor $R^a_{bcd}$, the Ricci curvature tensor $R_{ab}$, the Ricci scalar $R = g^{ab}R_{ab}$, and the determinant of the metric $|g|$. In terms of these objects the Einstein-Hilbert action is
\beq
\CS = \frac{1}{2\kappa^2} \int d^{D+1} x \sqrt{|g|} \(R - 2 \Lambda\)
\eeq
where $\Lambda < 0$ is the cosmological constant and the coupling constant is $\kappa^2 = 8 \pi G_N$ in terms of Newton's constant. The equations of motion following from this action are Einstein's equations in the absence of matter,
\beq
R_{ab} - \frac{1}{2} R g_{ab} + \Lambda g_{ab} = 0.
\eeq
Anti-de Sitter space with metric
\beq
ds^2 = \frac{L^2}{r^2} ( - dt^2 + dr^2 + d\vec{x}^2 )
\eeq
solves the source-free Einstein equations provided $\Lambda  = -f_d/L^2$ with $f_d$ a constant.

According to the holographic dictionary, this AdS spacetime is dual to the ground state of a conformal field theory whose UV we may think of as residing at $r=0$ (the boundary of space). With the above coordinate system the $x^\mu = (t,\vec{x})$ can be identified with coordinates in the boundary CFT while $r$ is a coordinate for the emergent radial dimension. The meaning of the duality is partially elucidated by matching symmetries. Poincare transformations on the CFT are represented on the bulk coordinates as $r \rightarrow r$ and $x^\mu \rightarrow A^\mu_\nu x^\nu + b^\mu$. This transformation is an isometry of the AdS metric. Similarly, scale transformations in the CFT are represented on the bulk coordinates as $r \rightarrow \lambda r$ and $x^\mu \rightarrow \lambda x^\mu$ which is again an isometry of the AdS metric.

If the CFT ground state is dual to emtpy AdS then more general spacetimes are dual to excited states of the CFT. Note, however, that not every CFT state has a nice geometrical description; only special semi-classical states will be dual smooth geometries in the bulk. The conditions on the CFT for the existence of a nice gravitational dual are still not fully understood, but a partial discussion may be found in the above reviews.

Finally, we must specify how entropy in the boundary CFT is to be encoded in the bulk. The answer is given by the Ryu-Takayanagi prescription which works as follows. Fix a boundary region $A$ whose entropy we wish to compute. Then consider the set $\Sigma$ of all bulk surfaces $\sigma$ which have $\partial \sigma = \partial A$ and which are homologous to $A$ (meaning the closed surface obtained by gluing $A$ and $\sigma$ along their common boundary is contractible). The entropy of region $A$ is then
\beq
S(A) = \min_{\sigma \in \Sigma} \frac{|\sigma|}{4 G_N}
\eeq
or the area of the minimal surface in Planck units. This formula,
which applies to Einstein gravity, generalizes the Bekenstein-Hawking formula for black hole entropy and is supported by a great deal of evidence.

{\bf Conditional mutual information for black hole states.} Now we turn to an actual calculation of the conditional mutual information in a finite temperature state of the CFT described by a bulk black hole. The black hole geometry is
\beq
ds^2 = \frac{L^2}{r^2} \( - f(r) dt^2 + \frac{1}{f(r)} dr^2 + d\vec{x}^2 \)
\eeq
where $f(r) = 1 - (r/r_h)^{d+1}$ and $r_h$ is the location of the event horizon. The horizon position $r_h$ is related to the temperature by $T r_h = \frac{d+1}{4\pi}$. One can easily verify that the above metric again solves Einstein's equations with the same negative cosmological constant as for empty AdS.

Let us begin with the case of $d=1$. The length of a minimal geodesic associated to a boundary length $\ell$ is proportional to the entropy formula written above,
\beq
|\sigma_\ell| \propto \log \sinh( \pi T \ell).
\eeq
To review how the minimal area formula works, consider now two intervals of length $\ell$ separated by a distance $x$; the intervals may be taken to be $[0,\ell$ and $[\ell+x,2\ell+x]$. There are two competing minimal surfaces, $\sigma_1$ which connects $x=0$ to $x=\ell$ and $\ell+x$ to $2\ell+x$ and $\sigma_2$ which connects $0$ to $2\ell +x$ and $\ell$ to $\ell+x$. The curve of minimum length for a given $x,\ell$ determines the entropy of the pair of intervals. There will be a critical $x$ such that $\sigma_2$ dominates for $x < x_c$ while $\sigma_1$ dominates for $x > x_c$. The curves have equal length at $x=x_c$ so $x_c$ is determined by
\beq
\sinh(\pi T (2\ell+x)) \sinh(\pi T x) = \sinh^2(\pi T \ell)
\eeq
to be $x_c = \frac{\log 2}{2\pi T}$ in the limit $\ell \gg 1/T$. This computation shows that the mutual information between the two intervals is large for $x<x_c$ and then drops to zero at $x=x_c$ and remains zero thereafter.
The suddenness of the transition is an artifact of the large-$N$ limit
which allows the classical gravity approximation.

Now what about the actual computation of interest, the conditional mutual information $I(A:C|B)$? $A$ is a set of intervals each of length $\ell_0$, $B$ is a set of intervals surrounding $A$ each of length $\ell_b$, and $C$ is a set of intervals each of length $\ell_1$. The four entropies needed are $S(AB)$, $S(BC)$, $S(B)$, and $S(ABC)$. $S(ABC)$ is just the total thermal entropy (which must be regulated in the planar limit we consider but is just proportional to total system size). Following the two interval calculation above, it may be verified that if $\ell_{1,b,2} \gg 1/T$ then the dominant geodesic is the one which which connects each interval to itself. Then the conditional information per unit cell (each unit cell contains one interval from $A$, two from $B$, and one from $C$) is
\beq
\frac{I(A:C|B)}{\text{unit cell}} = \frac{c}{3}\log\(\frac{\sinh(\pi T (\ell_0+2\ell_b)) \sinh(\pi T (\ell_1+2\ell_b))}{\sinh^2(\pi T \ell_b) e^{\pi T (\ell_0 + \ell_1 + 2\ell_b)}}\),
\eeq
where $c$ is the central charge (the large-$N$ limit being a large-$c$ limit here). In the limit $\ell_{0,b,1} \gg 1/T$ the above formula reduces to
\beq
\frac{I(A:C|B)}{\text{unit cell}} \sim c e^{-\ell/\xi},
\eeq
so taking $\ell = \xi \log\(\frac{c N_{\text{cells}}}{\epsilon}\)$ suffices for small reconstruction error.

Note that the holographic limit of large $c$ causes a further blowup in $\ell$, growing like $ \log(c)$. This is a relatively mild blowup (\eg,~$c=10^{10}$ only increases $\ell$ by a factor of $10$) but it is likely suboptimal. We conjecture that this blowup is artificial in the sense that there is a quasi-local channel of range $\xi$ which exactly produces the thermal state. To understand why the factor of $c$ appears, consider $c$ copies of a free fermion CFT. As we showed above the free fermion system has a quasi-local channel of range $\xi$ which exactly reproduces the thermal state, but if we wish to truncate this channel to a strictly local channel and maintain small total error in trace norm, say, then this will require a range that grows like $\xi \log c$ because each copy must be accurate to error $\epsilon/c$ so that the total error over all $c$ copies is $\epsilon$. Hence the $c$ arises from the very stringent notion of approximation employed.

Finally, while the geometric computation gives the leading order in large $c$ answer, there are also subleading in $c$ terms and the $\CO(1)$ term must be analyzed since even an order one term in $\CI$ can obstruct the use of \cite{fr1} and \cite{fr2}. Now we know in $d=1$ from the general CFT computation that the $\CO(1)$ term is also exponentially small; for completeness let us recall how this result is obtained holographically  \cite{Faulkner:2013ana}. $1/c$ corrections correspond to quantum corrections in the bulk gravity theory and the first quantum correction to $S(A)$ is obtained by evaluating the entanglement entropy of the interior of $A\cup \sigma(A)$ in the bulk. That is, we regard the bulk quantum fields (metric, et cetera)
as living in a fixed curved background and evaluate their entanglement entropy in the bulk region bounded by $A$ and $\sigma(A)$. Since the background geometry is thermal and since the bulk fields are free to leading order, the bulk entanglement will have the same cancellation properties as the bulk area law entanglement.\footnote{More explicitly, since the bulk fields are free to leading order, their state and hence their entropy is determined just by their 2-point correlation functions. These correlation functions are related to boundary 2-point functions, and the exponential decay of the boundary 2-point function implies the exponential decay of the bulk 2-point function. An exponentially decaying 2-point function for a free field implies the desired entropy form.}

Finally, although we worked through the case of $d=1$ in detail, the story is general to any dimension. Because of the minimization involved in the large $N$ contribution to the entanglement and because the relevant minimal surfaces all hug the horizon when the linear size is much greater than $1/T$, it follows that \eqref{eq:entropyform} is obeyed. The first quantum correction can also be controlled by studying the entanglement entropy of bulk fields in the black hole geometry (which typically leads to decaying correlations). We defer a more complete analysis to \cite{Hubeny-Swingle}.

\section{RG decomposition of $s=0$ fixed points}

\label{sec:RGdecomposition}

While the examples considered above were ultimately $s=0$, the range of the channels/unitaries required grew with increasing correlation length, \eg~as $T\rightarrow 0$.
In fact, there is additional structure in the problem which may ameliorate this situation. Said differently, the Petz and Fawzi-Renner (FR) reconstruction results guarantee that the recovery map does not depend on the state
outside the buffer region \cite{fr2}, but do not guarantee that the promised channel is local on the buffer region itself.

The converse situation provides a useful illustration:
consider a classical thermal state of a spin system: $ \rho = e^{ -  H/T }/Z$, where
by classical we mean
$H$ is diagonal in a product basis.
Such a density matrix is exactly Markov ($I(A:B|C) =0 $ for any regions $A,B,C$),
but can have long-range correlations if, for example, we choose $ T$
to be the critical temperature for the classical spin model. The resulting reconstruction channel in this case
is certainly not local on $B$.

A random channel with the allowed support grossly overestimates
the mutual information between subsets of the buffer region
$I(R_1 \subset B, R_2 \subset B)$ and
their entanglement entropies
$ S(R \subset B)$.
It is natural to propose a refinement of the FR results which says
that if
the mutual information between subsets of the buffer region
$I(R_1 \subset B, R_2 \subset B)$ vanishes (or is small)
then a(n approximately) local recovery channel $\CR$ should exist.
Known forms of the recovery map (such as \eqref{eq:petz-map})
have this property that when $\rho_B$ is factorized, then $\CR$ is local.

Here we take an RG viewpoint and argue that the channels exhibited above can be further decomposed into shorter range channels which, roughly speaking, implement conventional zero temperature renormalization until the scale $1/T$ is reached. The basic picture is this: many of the examples considered above, \eg~CFTs, are $s=1$ at $T=0$. This is really still a conjecture, but the evidence (numerical and analytical) is almost overwhelming in favor of this conjecture.\footnote{For example, there are numerous numerical studies in one and two dimensions, free particle results, and controlled constructions for special CFTs and certain non-conformal but scale invariant critical points. Moreover, many results about entropies including newer results about the regulated Schmidt rank are consistent with the existence of such an RG decomposition. Finally, recent conjectures relating complexity and geometry in the context of AdS/CFT also support the existence of an RG decomposition.} At finite $T$ the $T=0$ RG structure is not immediately lost. A manifestation of these facts is that the range of all our channels diverge as $T \rightarrow 0$.

However, this divergence of the range is a false problem. On physical grounds what is happening is that the channels constructed above can be obtained from a product of fixed ranged channels and it is the number of terms in this product which is diverging as $T \rightarrow 0$. Nevertheless, because of the local nature of the fixed range channels and because the number of terms in the product of channels is bounded by $\log(\text{IR cutoff})$, calculation of local observables is still formally efficient.

Once more, the physical picture is that we have approximately the ground state circuit for scales small compared to $1/T$ and a crossover to the finite temperature physics beyond $1/T$. In particular, each coarse-graining raises the effective temperature so that when the correlation length reaches the lattice scale the state becomes completely short-ranged and the above discussion can be used to cap off the flow into a product state.

{\bf Finite temperature crossover from $s>0$ to $s=0$:} The RG structure can be illustrated using an analysis of the thermal entropy. Recall that the spatial dimension is $d$, the temperature is $T$, and the linear size is $L$. The UV energy scale is $\Lambda$ and the UV length is $a$. We consider a broad class of scale invariant theories described by two numbers, the dynamical exponent $z$ and the hyperscaling violation exponent $\theta$. In such theories the thermal correlation length is $\xi(T) \sim T^{-1/z}$ and the entropy is $S(L,T) = c L^d T^{\frac{d-\theta}{z}}$. CFTs have $z=1$ and $\theta=0$.

We previously argued
\cite{Swingle:2015ipa}
that $s = 2^\theta$ for the ground states of such theories; the above analysis implies that the finite temperature state has $s=0$, but the range of the required unitary is of order $\xi$. The scaling structure further implies that the range $\xi$ unitary can be decomposed into $\log(\xi/a)$ unitaries of fixed range (range independent of $T$). Let us now see how the ground state value of $s$ persists until roughly the thermal scale is reached.

We compare the entropy of a size $L$ system at temperature $T$ to the entropy of a size $L/2$ system at temperature $2^z T$ (thermal length $\xi( 2^z T ) = \xi(T)/2)$. The former is $S(L,T) = c L^d T^{\frac{d-\theta}{z}}$ while the latter is
\beq
S(L/2, 2^z T) = c L^d T^{\frac{d-\theta}{z}} \frac{(2^z)^{\frac{d-\theta}{z}}}{2^d} = c L^d T^{\frac{d-\theta}{z}} 2^{-\theta}.
\eeq
Thus the entropy obeys
\beq
S(L,T) = 2^\theta S(L/2, 2^z T),
\eeq
so the entropy of $s = 2^\theta$ copies at temperature $2^z T$ and size $L/2$ is equivalent to the entropy of a single copy at temperature $T$ and size $L$. The ground state value of $s$ emerges at finite temperature as the number of copies of the thermal state needed to match entropies after rescaling lengths by a factor of $2$.

As the system shrinks it also heats up, and once the temperature reaches the microscopic energy scale $\Lambda$ the thermal state will have a thermal length $\xi$ of order the microscopic cutoff $a$. The scaling theory no longer correctly describes such high temperatures, but now our general reconstruction results can be applied resulting in a unitary with microscopic range which maps the system to a product state. At this point the theory has fully crossed over to $s=0$.

In the above, in speaking about unitaries, what we have in mind is a channel or a unitary acting on the thermal double state. It is interesting to note, however, that in the ideal scaling limit the Renyi entropies
$S_\alpha \equiv {1\over 1 - \alpha} \tr \rho^\alpha$ also obey
\be\label{eq:renyis-scaling} S_\alpha(L,T) \buildrel{?}\over{=} 2^\theta S_\alpha(L/2,2^z T) . \ee
For a mathematical CFT $(z=1,\theta=0)$ without a cutoff, this is the statement that the density matrix
depends only on the product $LT$.
Renyi entropies with small $\alpha$ are sensitive to the physics
at the short-distance cutoff
and as such the scaling theory breaks down.
An exact equality of the form \eqref{eq:renyis-scaling} would imply that the two states, $L$ and $2^\theta$ copies of $L/2$, have the same spectrum, an interesting statement that may be approximately true.

Once more, the physical picture is this. Allowing a range $\xi$ unitary, it was shown that the thermal state is $s=0$ for all $z$ and $\theta$ provided it is adiabatically connected to the high temperature limit. Allowing only fixed (microscopic) range unitaries, the thermal state crosses over from $s = 2^\theta$ to $s=0$ as it is renormalized. On physical grounds such fixed-range unitaries should exist, but in most cases they have yet to be constructed. This construction is beyond the scope of this work, but is an exciting direction for further work. Some systems where progress has been made or can be made follow. Free-particle techniques should allow such a construction in those models
\cite{PhysRevB.81.235102, 2015WhiteFreeFermions, Evenbly:2016cly}.
There has been a recent proposal for a related construction for thermal states of $1+1$ CFTs using their special symmetry properties \cite{Czech:2015xna}. Numerical constructions can also be carried out in some cases. It would also be interesting to extend the analysis of Ref.~\cite{sqrt-draft} to finite temperature.

\section{Bubble-of-Nothing analysis of topological gauge theories}
\label{sec:gauge-theory}

In this section we first review the obstructions to the bubble-of-Nothing array argument
of \S\ref{sec:wormhole-argument-here}
presented by long-range entanglement at $T=0$. As representative examples, we discuss $p$-form gauge theories in $d$ spatial dimensions. These topological obstructions prevent long-range entangled states from being $s=0$ at $T=0$, but they are all $s=1$ states \cite{sourcery1}. Then we consider $T>0$ and show that $1$-form gauge theories become $s=0$ but $2$-form gauge theories in $d=4$ remain $s=1$ at small but non-zero temperature.
Because its topological order survives at $T>0$,
the latter system is a
 robust self-correcting quantum memory
\cite{Dennis:2001nw, 2007PhRvB..76r4442C, 2008arXiv0811.0033A,2011PhRvL.107u0501H, Mazac:2011np,Grover:2011fa, 2014PhRvL.112g0501H}.
Our analysis indicates that this is a general property of
 $s>0$ finite-temperature fixed points.

\subsection{Obstruction theory at $T=0$}

We focus on $Z_2$ $p$-form gauge theory in $d$ dimensions. The analysis can be extended to a wide variety of other kinds of gauge theories. Such a gauge theory is a topological quantum field theory which means it has a topologically protected ground state space on a manifold with non-trivial topology. Even on a manifold with trivial topology where there is a unique ground state, that ground state is long-range entangled and cannot be deformed to a product state using a finite depth quantum circuit.

One way to see that the state is non-trivial is to attempt to perform the cellular construction we used above for other short-range correlated states. Indeed, correlations are exponentially decaying in the ground state of a topological field theory, so one might have hoped that \eqref{eq:entropyform} was valid, but as we now discuss the appearance of certain topological terms in the entropy foil the validity of \eqref{eq:entropyform}.

According to \cite{Grover:2011fa}, the entanglement entropy of a region $A$ in a $p$-form gauge theory
(in the zero-correlation-length, solvable limit)
may be written as
\beq
S(A) = \text{area law terms} - \gamma(A)
\eeq
where the topological term is
\beq
\gamma(A) = \sum_{k=0}^{p-1} (-1)^{p-1+k}b_k(\partial A)
\eeq
and $b_k(\partial A)$ is the $k$-th Betti number of the boundary of $A$. For example, for $p=1$ the topological term is $\gamma(A) = b_0(\partial A)$, \eg~just the number of disconnected components in the boundary of $A$. We will use this data to demonstrate obstructions to reconstructing the state using the bubble-of-Nothing array argument.

Our cellular reconstruction procedure is defined using a hypercubic lattice. We begin with $0$-cells, the vertices of the hypercubic lattice. There is $1$ $0$-cell per unit cell of the lattice. Then come the $1$-cells, the links of the lattice. There are $d$ $1$-cells per unit cell. Going to the general case, there are $\frac{d!}{(d-k)!k!}$ $k$-cells per unit cell. We proceed as in the previous discussion by attaching $k$ cells to the $k-1$-skeleton (the union of all $q$-cells with $q<k$). At each step we begin with the $k-1$-skeleton and end up with the $k$-skeleton, but we must also keep track of the intermediate buffer zones during the reconstruction.

{\bf General argument for obstruction pattern.}
As we show via examples, the general pattern is that the $p$-form gauge theory has at most two obstructions, one occurs when going from the $p-1$-skeleton to the $p$-skeleton and one occurs when going from the $d-p-1$-skeleton to the $d-p$ skeleton. If $d = 2p$ then these two obstructions collapse to one. The obstructions are related by a duality transformation.
Morally, this happens because it is at these steps that
one gains the cycles on which to measure the charges of
the topological excitations by Gauss' law.
For example, for one-form gauge theory
(where the degrees of freedom are associated to the links)
in $d=2$ once we have complete loops we can detect the presence of fluxes. Whenever we would be able to detect a new kind of topological charge after a step of the bubble-of-Nothing growth procedure, there must be an obstruction to the step. In other words, there is an obstruction whenever we incorporate new topological data into the state.
We anticipate that this purely mathematical statement can be verified in general by a Mayer-Vietoris
argument on the cells but have not completed that analysis.

We now give a series of examples. The examples consist of calculations of the conditional mutual information per unit cell at each step of the cellular reconstruction. Obstructions occur when the conditional mutual information does not vanish. We focus exclusively on the topological terms since all conventional local area law terms will cancel from the conditional mutual information. Throughout, disks refer to $d$-disks or equivalently $d$-dimensional balls.

Example 1 ($d=2$, $p=1$):
\begin{itemize}
\item $0\rightarrow 1$: $A$ is a disk per unit cell, $B$ is four disks per unit cell, $C$ is two disks per unit cell, $AB$ is a disk per unit cell, $BC$ is two disks per unit cell, and $ABC$ is the $1$-skeleton. The entropies per unit cell are $S(AB)=-1$, $S(BC)=-2$, $S(B)=-4$, and $S(ABC)=-1$. The conditional mutual information is $I(A:C|B) = 2$ and there is an obstruction.
\item $1\rightarrow 2$: $A$ is the 1-skeleton, $B$ is an annulus per unit cell, $C$ is a disk per unit cell, $AB$ is the $1$-skeleton, $BC$ is a disk per unit cell, and $ABC$ is the $2$-skeleton, which is fills 2d space. The entropies per unit cell are $S(AB)=-1$, $S(BC)=-1$, $S(B)=-2$, and $S(ABC)=0$. The conditional mutual information is $I(A:C|B) = 0$ and there is no obstruction.
\end{itemize}

Example 2 ($d=3$, $p=1$):
\begin{itemize}
\item $0\rightarrow 1$: $A$ is a disk per unit cell, $B$ is six disks per unit cell, $C$ is three disks per unit cell, $AB$ is a disk per unit cell, $BC$ is three disks per unit cell, and $ABC$ is the $1$-skeleton. The entropies per unit cell are $S(AB)=-1$, $S(BC)=-3$, $S(B)=-6$, and $S(ABC)\approx 0$.
Here is the meaning of $\approx$: The $1$-skeleton indeed has a boundary, but it has one boundary for the whole sample,
so the betti-number per unit cell vanishes in the thermodynamic limit.

The conditional mutual information is $I(A:C|B) = 2$ and there is an obstruction.
\item $1\rightarrow 2$: $A$ is the 1-skeleton, $B$ is three donuts per unit cell, $C$ is three disks per unit cell, $AB$ is the $1$-skeleton, $BC$ is three disks per unit cell, and $ABC$ is the $2$-skeleton. The entropies per unit cell are $S(AB)\approx 0$, $S(BC)=-3$, $S(B)=-3$, and $S(ABC)=-1$. The conditional mutual information is $I(A:C|B) = 1$ and there is an obstruction.
\item $2 \rightarrow 3$: $A$ is the 2-skeleton, $B$ is a shell per unit cell, $C$ is a disk per unit cell, $AB$ is the $1$-skeleton, $BC$ is a disk per unit cell, and $ABC$ is the $3$-skeleton. The entropies per unit cell are $S(AB)=-1$, $S(BC)=-1$, $S(B)=-2$, and $S(ABC)=0$. The conditional mutual information is $I(A:C|B) = 0$ and there is no obstruction.
\end{itemize}

Example 3 ($d=4$, $p=2$):  For $p=2$, $\gamma(A) = b_0(\partial A) - b_1(\partial A)$.
Details of this calculation are organized into a table in the appendix \ref{appendix:table}.
We must confess that we have had to infer the value of
one of the Betti numbers in this calculation ($b_0(\partial \Sigma_2)$)
from our ansatz; the consistency of rest of the calculation
is still a strong check of the ansatz.
\begin{itemize}
\item $0\rightarrow 1$: $A$ is a 4-disk per unit cell, $B$ is 8 4-disks per unit cell, $C$ is 4 disks per unit cell, 
and $ABC$ is the $1$-skeleton, $\Sigma_1$.
The boundaries of the disks are simply connected
and the boundary of $\Sigma_1$ is a 4-dimensional generalization
of the `plumber's nightmare', which has $b_1( \partial \Sigma_1) \simeq 0$ per unit cell
and $b_1( \partial \Sigma_1) = d-1 = 3$.
The entropies per unit cell are $S(AB)=1$, $S(BC)=4$, $S(B)=8$, and $S(ABC) = 3$.
The conditional mutual information is $I_{0\to1}^{p=2}(A:C|B) = 3-3=0$ and there is no obstruction in $p=2$ gauge theory.
(In $p=1$ gauge theory, the obstruction at this stage would be $ I_{0\to1}^{p=1}(A:C|B)  = 3$.)

\item $1\rightarrow 2$: $A$ is the 1-skeleton, $B$ is 6 donuts per unit cell
(whose boundaries are each $S^2 \times S^1$), $C$ is 6 disks per unit cell,
and $ABC$ is the $2$-skeleton.
The entropies per unit cell are $S(AB) =b_0 - b_1 = -3 $, $S(BC)=6$, $S(B)=6-6=0$,
and $S(ABC)=-4$. The conditional mutual information is $I_{1\to 2}^{p=2}(A:C|B) \simeq 7$ and there is an obstruction.
(In $p=1$ gauge theory there would be no obstruction here $I_{1\to 2}^{p=1}(A:C|B)   = 0$.)

\item $2\rightarrow 3$: $A$ is the 2-skeleton, $B$ arises
by intersecting the 3-volumes with the faces and is
4 thickened 4-shells per unit cell,
$C$ is 4 disks per unit cell,
and $ABC$ is the $3$-skeleton, $\Sigma_3$, which
is everything minus the 4-volume filling.
The obstruction for $p=1$ gauge theory would be $I_{2\to 3}^{p=1}(A:C|B)= - b_0(\partial \Sigma_3) = -1$.
For $p=2$, $I_{2\to3}(A:C|B) =
0 $ and there is no obstruction.
(In $p=1$ gauge theory there would be an obstruction here, $I_{2\to 3}^{p=1}(A:C|B)   = 1$.)

\item $3 \rightarrow 4$: $A$ is the 3-skeleton, $B$ is a shell per unit cell, $C$ is a disk per unit cell, $AB$ is the $3$-skeleton, $BC$ is a disk per unit cell, and $ABC$ is the whole space.
The boundaries are all simply connected in this case, so
the entropies per unit cell are $S(AB)=1$, $S(BC)=1$, $S(B)=2$, and $S(ABC)=0$. The conditional mutual information is $I(A:C|B) = 0$ and there is no obstruction.
\end{itemize}

\subsection{Fate of obstructions at $T>0$}

\label{sec:long-range-entanglement-at-finite-T}

The obstructions to
an $s=0$ reconstruction
described above arise from long-ranged entanglement
in the groundstate.
Such topological order is destroyed by the proliferation
of defects of the appropriate nature.
For 1-form gauge theory, the appropriate defects
are particles;
the topologically-protected degenerate groundstates
differ by the action of the holonomy of these particles
around topologically non-trivial curves (Wilson loop operators).
Any finite temperature introduces a system-size-independent
density of these particles, $n(T)\propto e^{ - \Delta/T}$,
where $\Delta$ is the energy gap.
The obstruction persists only for regions much smaller than
the average spacing between these particles, $ R \ll l_q(T) \equiv {1\over n(T)^{1/d} }$
\cite{2007PhRvB..76r4442C}.
This is a nice illustration of the discussion of \S\ref{sec:RGdecomposition}:
for lengths smaller than the inter-particle spacing $ l_q(T)$,
the system is $s=1$,
while for longer lengths, it is $s=0$.

The situation for $(p\geq 2)$-form gauge theory in $d>3$ is different \cite{Dennis:2001nw, 2007PhRvB..76r4442C, 2008arXiv0811.0033A,2011PhRvL.107u0501H,
Mazac:2011np,Grover:2011fa, 2014PhRvL.112g0501H}.
($d> 3$ is required because $2$-form gauge theory in $d=3$ dimensions
can be dualized to $1$-form gauge theory.)
The defects which destroy the topological order are large, closed strings
(more generally, $p-1$-dimensional objects).
On a generic space, they must be large
(scaling with system size) because
they must wrap
topologically non-trivial one-cycles.
The Bolzmann factor therefore
provides a system-size-dependent suppression of the density of such defects.

The statistical mechanics of strings whose energy is dominated by a tension term
is governed by a Hagedorn equation of state (\eg~\cite{Atick:1988si, Horowitz:1996nw, Horowitz:1997jc}):
the entropy at fixed energy is linear in the energy, $S(E)  = a E$,
with a coefficient determined by the string tension.
Therefore, the free energy $ F = E - T S = (a -  T ) E $.
This Peierls-type argument implies a transition at some `Hagedorn' temperature
above which the canonical ensemble in terms of strings breaks down.
Above this temperature, strings are condensed,
and in the 2-form gauge theory context,
the topological order is destroyed.
Conversely, below the Hagedorn temperature,
the entropic contribution is overwhelmed by the tension,
and the ensemble is dominated by small strings.
Hence, there is a temperature below which
the 2-form topological order persists.
Evidence for such a finite-temperature transition
in 4d 2-form gauge theory
has been found in numerical work \cite{2014PhRvL.112g0501H}.

%

\section{Discussion, conjectures and questions}
\label{sec:open-questions}

In this paper we defined a new notion of mixed s-sourcery which generalizes our previous pure state construction to mixed states. We showed that a huge variety of finite temperature states of matter, including free particles (\S\ref{sec:free-fermions}), conformal field theories (\S\ref{sec:CFT} ), topological phases (\S\ref{sec:gauge-theory}), and holographic states (\S\ref{sec:holography}) fall into our scheme. We also argued for a further renormalization-group inspired decomposition of the local unitaries involved in the s-sourcery construction (\S\ref{sec:RGdecomposition}).

A major theme of our work is the idea that thermal double states, which are purifications of thermal mixed states, can be cast as unique ground states of local Hamiltonians. To give a general argument for this conclusion in models with the right entropic properties, we used the rapidly advancing machinery of approximate quantum conditional independence. This technology is quite general, so we expect that it can be applied much more widely.

In this final discussion section, we make some comments on issues raised by our work. We also discuss a few applications of this work and mention some open questions.

\subsection{Form of local channel arising from approximate conditional independence}

In our calculations above, we focused on the information theoretic conditions for approximate conditional independence, namely the near vanishing of the conditional mutual information $I(A:C|B) = S(AB) + S(BC) - S(B) - S(ABC)$. We did not give an explicit formula for the channel which glued the system back together, except in special cases, \eg~free particle models. As reviewed in \S\ref{subsec:qmarkov}, such a general formula does exist in the case when $I(A:C|B)$ exactly vanishes, but this formula is sufficiently complex that further work is needed to concretely apply it to our problem. Furthermore, in most of the situations we considered, the conditional mutual information did not exactly vanish.

Recently there has been considerable progress in exhibiting explicit reconstruction maps
for the case of approximate conditional independence
\cite{2015RSPSA.47150338W, 2015arXiv150907127J}. Here we quickly review these works and explain how they may be used to construct the channels we need. We also give a procedure to construct thermal double Hamiltonians using the given channels.

In \cite{2015arXiv150907127J} the recovery problem is framed in a very general setting. Consider two quantum states $\rho$ and $\sigma$ and a channel $\CN$. The condition for approximate recoverability is phrased in terms of the relative entropy, $D(\rho \| \sigma) = \text{tr}(\rho \log \rho - \rho \log \sigma)$. If $D(\rho \| \sigma) - D(\CN(\rho) \| \CN(\sigma) ) \approx 0$, then there exists a channel $\CR_{\CN,\sigma} $ such that $\CR_{\CN,\sigma}(\rho) \approx \rho$. The precise theorem is
\beq
D(\rho \| \sigma) - D(\CN(\rho) \| \CN(\sigma) ) \geq -2 \log F[\rho , \CR_{\CN,\sigma}(\CN(\rho))]
\eeq
where $F(\rho,\sigma) = \|\sqrt{\rho}\sqrt{\sigma}\|_1$ is the fidelity and where $\CR_{\CN,\sigma}$ is defined by
\beq \label{eq:recoverymap}
\CR_{\CN,\sigma}(X) = \int_{-\infty}^\infty dt \, \frac{\pi}{\cosh (2 \pi t) + 1} \sigma^{1/2 - it} \CN^\dagger \left[ \CN(\sigma)^{-1/2+it} X \CN(\sigma)^{-1/2-it}\right] \sigma^{1/2+it}.
\eeq

To apply this formalism to the case of three regions $A$, $B$, and $C$ with $I(A:C|B) \approx 0$, let $\CN = \text{tr}_A$ and let $\rho = \rho_{ABC}$ and $\sigma = \rho_{AB} \otimes \rho_C$. Now we compute
\beq
D(\rho\|\sigma) - D(\CN(\rho)\| \CN(\sigma)) = [ - S(ABC) + S(AB) + S(C)] - [ - S(BC) + S(C) + S(B)].
\eeq
But this expression, upon cancelling the $S(C)$ factors, is $I(A:C|B)$. Assuming $I(A:C|B) \approx 0 $, then \eqref{eq:recoverymap} defines an explicit recovery channel which undoes the action of $\text{tr}_A$ on $\rho_{ABC}$. Furthermore, note that $\sigma = \rho_{AB} \otimes \rho_C$ depends only on the marginals $\rho_{AB}$ and $\rho_C$ and not on the full state $\rho_{ABC}$. Hence $\CR_{\CN,\sigma}$ also depends only on the marginals and not on the full state. Since $\CN(\rho_{ABC}) = \rho_{BC}$, we have that
\beq
\rho_{ABC} \approx \CR_{\CN,\sigma}(\rho_{BC})
\eeq
as desired.

Our demonstration of approximate conditional independence thus shows that we can in effect ``untrace out" the system starting from nothing using a series of steps corresponding to our cellular decomposition of space. Furthermore, given the composite local channel which untraces out the whole system, we may construct a local Hamiltonian for a thermal double as follows. First, resolve the untracing out channel into a Kraus representation. Second, introduce an environment to realize the Kraus operators as the action of a unitary transformation. Third, by conjugating with the resulting unitary transformation, we may transform a trivial Hamiltonian whose exact ground state is the initial product state into a Hamiltonian whose ground state is a thermal double state. It will be extremely interesting in the future to carry out these steps for a variety of models.

\subsection{Thermal double Hamiltonians}

For a wide variety of physical systems, the preceding arguments show the existence of a local, hermitian Hamiltonian $H_T$ whose ground state is approximately the thermal double state $\ket{T}$. However, the range of this Hamiltonian in general has to grow as $T$ decreased because the range is of order the thermal length $\xi$. Moreover, the terms in the Hamiltonian can be very many-body: at this level of generality we can only say that each term must contain $\sim \xi^d$ or fewer operators.

Although this represents the best that our general information theoretic arguments can do, with additional physical input we can say more. For example, in the case of free particles we constructed an $H_T$ consisting only of two-body terms of range $\xi$. It is interesting to investigate more generally under what conditions a long-ranged but few-body $H_T$ exists. For example, given the thermal state of a CFT, we could ask for several refinements, \eg~(1) an $H_T$ which is strictly short-ranged but does not have an energy gap or (2) an $H_T$ with an energy gap which is long-ranged but few-body.

{\bf Simple example:} These questions are largely open, but here we give one example where progress is possible. Consider $N$ spin-1/2 degrees of freedom with Pauli matrices $X_i$, $Y_i$, and $Z_i$. A classical thermal state is a state of the form
\beq
\rho(T) = \sum_s \frac{e^{- h(s)/T}}{Z} \ket{s} \bra{s}
\eeq
where $s$ labels the spin configuration, $Z_i \ket{s} = s_i \ket{s}$, and $h(s)$ is a classical energy. This special class of thermal states is generally interesting as the high temperature limit of thermal states of quantum spin systems.

A simple thermal double state for $\rho(T)$ is
\beq
\ket{T} = \sum_s \sqrt{\frac{e^{- h(s)/T}}{Z}} \ket{s} \ket{s}.
\eeq
Remarkably, if we define the isometry $W : \CH \rightarrow \CH \otimes \CH$ by
\beq
W\ket{s} = \ket{s} \ket{s},
\eeq
then we see that
\beq
\ket{T} = W \ket{\psi}
\eeq
where
\beq
\ket{\psi} = \sum_s \sqrt{\frac{e^{- h(s)/T}}{Z}} \ket{s}
\eeq
is a ``square-root state" recently studied in \cite{sqrt-draft}.

There it was shown how to construct a local Hamiltonian for which $\ket{\psi}$ is the exact ground state (see also \cite{2012LNP...843..245A}). When the state $\rho(T)$ is short-range correlated it was shown that $\ket{\psi}$ is $s=0$, and when $\rho(T)$ corresponds to a classical critical point it was shown that $\ket{\psi}$ is $s=1$. These results immediately imply that $\rho(T)$ is a purified $s=1$ fixed point.

Let $H$ denote a positive local Hamiltonian with $H \ket{\psi} = 0$. Then $H_1 = W H W^\dagger$ is a positive local Hamiltonian with $H_1 \ket{T} = W H W^\dagger W \ket{\psi} = 0$. However, $\ket{T}$ is not the only zero-energy state: any state of the form $\ket{s} \ket{s'}$ with $s \neq s'$ is also zero-energy. To remedy this, define
\beq
H_2 = \sum_i \frac{I - Z_i \tilde{Z}_i}{2}
\eeq
where $\tilde{Z}_i$ refers to the second copy in the thermal double. $H_2$ punishes states $\ket{s} \ket{s'}$ with $s \neq s'$, so the total Hamiltonian
\beq
H_T = H_1 + H_2
\eeq
has as its unique ground state the thermal double state $\ket{T}$. We can similarly construct an RG circuit for $\ket{T}$ from an RG circuit for $\ket{\psi}$ using the isometry $W$.

The remarkable thing about these results is that the thermal double Hamiltonian is local even when $\rho(T)$ has long-range correlations! Of course, the thermal double Hamiltonian does not have an energy gap when the correlations are long-ranged. While the classical states considered here are clearly very special, they at least demonstrate that a strictly local few body thermal double Hamiltonian can sometimes exist. We speculate that such local Hamiltonians exist more broadly; after all, the two copies of the ground state (a zero temperature thermal double) are by assumption the ground state of a local Hamiltonian and this is in some sense the hardest case.

\subsection{Some further applications}

Here we give two additional applications of our work, one constructing a local dynamical evolution with the thermal as its fixed point and one upper bounding the complexity of thermal double states. These by no means exhaust the applications of our results.

{\bf Thermal states as fixed points of local Lindblad evolution.}
Using our results on approximate condition independence and the existence of an approximate recovery map, we show that suitable thermal states of local Hamiltonians can be cast as fixed points of an open system evolution equation, a Linblad equation.

Consider a $d$ dimensional disk $D_x$ centered at position $x$ with radius a few thermal lengths. Let $\CR_x$ be the recovery map obtained from the trace map $\tr_{D_x}$ as discussed above. Assuming the thermal state obeys \eqref{eq:entropyform}, then $\CR_x$ can be instantiated as a local map acting in a neighborhood of $D_x$.

$\CR_x$ maps the total system minus the disk to the total system, so it is convenient to define a new map from the total system to the total system. It is simply the composition of $\CR_x$ and $\tr_{D_x}$,
\beq
\Phi_x(\sigma) = \CR_x(\tr_{D_x}(\sigma)).
\eeq
$\Phi_x$ has two crucial properites: it obeys $\Phi_x(\rho(T)) \approx \rho(T)$ and it is local.

We wish to turn this into a dynamical map, that is into a rule for open system evolution. The Lindblad equation describes general Markovian open system evolution; it is written as
\beq
\partial_t \sigma = \CL (\sigma)
\eeq
where $\CL$ is the Lindblad superoperator. Consider a Lindblad superoperator defined by
\beq
\CL = \sum_x \( \Phi_x - \text{Id} \).
\eeq
Then we compute
\beq
\partial_t \rho(T) = \CL(\rho(T)) = \sum_x \(\Phi_x(\rho(T)) - \rho(T)\) \approx 0,
\eeq
so $\rho(T)$ is an approximate fixed point of the flow generated by $\CL$. With more work it should be possible to show that $\rho(T)$ is the unique fixed point and perhaps even bound the mixing time, see \cite{2014arXiv1409.3435K}.

{\bf Bounds on thermal double state complexity.}  Using the existence of the recovery channel, we can provide an upper bound on the state complexity of the thermal double state. Recall that the state complexity is defined as the minimum number of gates from a universal set needed to produce the state of interest from a reference state. Since we have a procedure for producing a thermal double state, we can give an upper bound on its complexity.

To get the bound, suppose that the unitary which purifies the action of the recovery channel is composed of blocks that act on $\ell^d$ degrees of freedom at a time. To have small error, $\ell$ should be taken to be of order $\xi \log L$ where $\xi$ is the thermal length and $L$ is the system size. The number of blocks is of order $ a (L/\ell)^d$. The maximum complexity of a block acting on $\ell^d$ degrees of freedom is of order $e^{b \ell^d}$ for some constant $b$ that depends on the gate set, the nature of the degrees of freedom, and so forth. Hence the total complexity obeys
\beq
\text{complexity} \leq a \left( \frac{L}{\ell}\right)^d e^{b \ell^d}.
\eeq
In $d=1$ this expression is polynomial in $L$, while for $d>1$ it is quasi-polynomial in $L$, \ie~the $\log$  is a polynomial in $\log L$.

By using the conjectured RG structure of the channel (\S\ref{sec:RGdecomposition}) and a more refined notion of complexity which counts gates weighted by their strength
\cite{2006Sci...311.1133N, Brown:2015lvg}, we can produce a more refined estimate of the complexity. Suppose that at each step of the RG, \eg~going from size $2^j a$ to size $2^{j-1} a$ ($a$ is a lattice spacing), the RG circuit is generated by a quasi-local unitary $U_j = e^{i K_j}$. The number of RG steps is $n_{RG} \sim \log_2(\xi/a)$ where $\xi$ is the thermal length; starting from linear size $L$, after $n_{RG}$ steps the system has size $L 2^{-n_{RG}}$ and all correlations are ultralocal.

For concreteness, consider a system of qubits arranged in $d$-dimensional array. Suppose that $K = \sum_{x,r} K_{x,r}$ where $K_{x,r}$ consists of $m_r$ terms supported on a disk of radius $r$ centered at $x$ and each term is proportional to a product of Pauli operators and has norm bounded by $f(r)/m_r$ for function $f(r)$ decreasing faster than any power of $r$ at large $r$. Measuring the complexity of $U_j$ by the number and size (\eg~the norm) of the terms in $K_j$ gives the following estimate:
\beq
\text{complexity}(U_j) \leq \sum_{x,r} \underbrace{m_r}_{\text{number}} \underbrace{\left(\frac{f(r)}{m_r} \right)}_{\text{size}} = \sum_{x,r} f(r) \leq \left( \frac{2^{j} a}{a}\right)^d f
\eeq
where $f \equiv \sum_r f(r) < \infty$. Writing $L = 2^J a$, the total complexity is then bounded by
\beq
\text{complexity} \leq \sum_{j=J-n_{RG}}^{J} 2^{j d} f = f \left[ \frac{2^{d(J+1)} }{2^d - 1} - \frac{2^{d(J-n_{RG})} }{2^d - 1} \right].
\eeq
In terms of $L$, $a$, and $\xi$ this is
\beq
\text{complexity} \leq \frac{f}{2^d - 1} \left[2^d \left(\frac{L}{a}\right)^d - \left(\frac{L}{\xi}\right)^d \right].
\eeq
Because $f$ may depend on temperature, this estimate does \emph{not} imply that the complexity of the thermal double is monotonically decreasing with increasing $T$.

\subsection{Conjectures and questions}

\begin{itemize}
\item
Especially given the connection to finite-temperature quantum memory,
it would be very interesting to find states which are mixed $s>0$.
Such a state must necessarily be somewhat exotic.
In the groundstate $s$-sourcery paper, Haah's cubic
code \cite{2011PhRvA..83d2330H} played a starring role as an example with $s=2$ \cite{2014PhRvB..89g5119H}.
Hence it is natural to ask whether or not Haah's code \cite{2011PhRvA..83d2330H} is mixed $s=0$ at finite $T$.
Haah's code is like $p>1$-form gauge theory (in $d>3$) in that
the defects which mix the topological sectors are supported on a locus of dimension larger than zero
(though less than one).
In fact, the energy barrier between sectors grows logarithmically with system size \cite{2011PhRvA..83d2330H, PhysRevLett.107.150504, 2013PhDHaah}.
However, at any finite temperature, the {\it multiplicity} of such defects also grows rapidly with system size,
and the entropic gain favors a proliferation of the defects, resulting in a
system-size-independent memory lifetime of order $ e^{ 1/T^2}$ \cite{PhysRevLett.111.200501}.
This suggests that the cubic code at any finite $T$ is adiabatically connected to $T=\infty$
and hence satisfies the hypotheses of \S\ref{sec:CFT}.
Very recently \cite{2016arXiv160307805S} used other techniques to show that
Haah's code is topologically trivially in Hastings' sense, and hence indeed mixed $s=0$.
A set of phases with similar properties, in that the
defects are not particles which are free to move everywhere in the system,
are the `higher-spin spin liquids' of
\cite{2016arXiv160405329P, 2016arXiv160608857P}.
Are these mixed $s=0$?
Another interesting class of examples to consider are
local Hamiltonians whose groundstates violate the area law
(these examples are in one dimension)
\cite{2012PhRvL.109t7202B, 2014arXiv1408.1657M, 2016arXiv160503842S}.

\item We conjecture that any two thermal states in the same phase can be related by a local channel. So far we proved this for free fermion states. Is the thermal double construction useful for answering this  conjecture? A problem is that even if there is family of gapped thermal double Hamiltonians, this does not imply the channel statement since the starting state (some particular thermal double state) need not factorize between system and environment as required in the channel definition.
\item Under what conditions is the thermal double the gapped ground state of a local Hamiltonian? Is the decay of correlations enough to guarantee the existence of such a local parent Hamiltonian?

\end{itemize}

{\bf Acknowledgements.}
We thank Claudio Chamon and Isaac Kim for helpful correspondence.
This work was supported in part by
funds provided by the U.S. Department of Energy
(D.O.E.) under cooperative research agreement
DE-SC0009919.

\appendix
\renewcommand{\theequation}{\Alph{section}.\arabic{equation}}

\section{Some relevant quantum information theory}
\label{sec:appendix}
\subsection{Entropy bounds}

Consider a mixed state $\rho = \sum_i p_i \ket{i}\bra{i}$ and let $\ket{\sqrt{\rho}} = \sum_i \sqrt{p_i} \ket{i}_1 \ket{\tilde{i}}_2$ be a purification of $\rho$ with the property that $\ket{\sqrt{\rho}}$ is an eigenstate of the swap operator which exchanges the two systems. Let us further suppose that the purified system decomposes as $A A^c \tilde{A} \tilde{A}^c$ where $A$ and $A^c$ are a bipartition of the first system and similarly for the tilde variables.

Then we have the following bound:
\beq
 I(A,A^c) \leq S(A\tilde{A}).
\eeq
This bound may be interpreted as saying that the combined entropy of $A$ and its corresponding region in the purifying second system bounds the mutual information in the original mixed state.

The lower bound is a consequence of strong subadditivity, purity, and the swap invariance of the two systems. We use overall purity to exchange various systems for their complements,
\beq
I(A,A^c) = S(A) + S(A \tilde{A}\tilde{A}^c) - S(\tilde{A} \tilde{A}^c).
\eeq
Strong subadditivity is used to write
\beq
S(A\tilde{A}\tilde{A}^c) \leq S(A\tilde{A}) + S(\tilde{A}\tilde{A}^c) - S(\tilde{A}).
\eeq
Combining these two formulas leads to the desired bound,
\beq
I(A,A^c) \leq S(A) - S(\tilde{A}\tilde{A}^c) + S(A\tilde{A}) + S(\tilde{A}\tilde{A}^c) - S(\tilde{A}) = S(A\tilde{A}),
\eeq
where in the last step we used $S(A) = S(\tilde{A})$.

One might also hope for an upper bound of some type, but there seem to be some some obstacles to obtaining a useful bound (due to the freedom to act with an arbitrary unitary on the purifying system). On the other hand, if the thermal double state can be chosen to be the ground state of a local Hamiltonian with an energy gap then the entanglement entropy is almost certainly bounded by an area law.

\subsection{Review of reversibility}
\label{subsec:reverse}

Given a quantum channel $\mathcal{E}$ and a state $\sigma$, we can always construct another channel $\mathcal{R}_{\mathcal{E},\sigma}$ such that
\beq
\mathcal{R}_{\mathcal{E},\sigma}(\mathcal{E}(\sigma)) = \sigma.
\eeq
If $\mathcal{E}$ has Kraus representation given by $\{ M_k\}$ then $\mathcal{R}$ has a Kraus representation given by \beq
\CR : \{ \sigma^{1/2} M_k^\dagger \mathcal{E}(\sigma)^{-1/2} \}.
\eeq
This channel is called the transpose channel or Petz channel \cite{petz1986, PETZ01031988, prettygoodchannel, 2003RvMaP..15...79P, qmarkov}. The cyclicity of the trace guarantees that the channel $\mathcal{R}$ is trace preserving,
\bea
\tr(\mathcal{R}(\rho)) && = \sum_k \tr\left(\sigma^{1/2} M_k^\dagger \mathcal{E}(\sigma)^{-1/2} \rho \mathcal{E}(\sigma)^{-1/2} M_k \sigma^{1/2}\right) \nonumber \\
&& = \sum_k \tr\left(\mathcal{E}(\sigma) \mathcal{E}(\sigma)^{-1/2} \rho \mathcal{E}(\sigma)^{-1/2} \right) = \tr(\rho).
\eea
Acting with $\mathcal{R}$ on $\mathcal{E}$ is also easily seen to reproduce the state $\sigma$ since the internal factors of $\mathcal{E}(\sigma)^{-1/2}$ cancel with the input.

Note that while $\mathcal{R}$ is defined with respect to $\mathcal{E}$ and a particular state $\sigma$, the hope, in an error correcting context say, is that $\mathcal{R}$ recovers not just $\sigma$ but also high probability pure states from the ensemble represented by $\sigma$.

For example, consider the toric code \cite{2003AnPhy.303....2K}
and suppose $\mathcal{E}$ is a dephasing channel (see \eqref{eq:dephasing} above) acting on a single link. Let $\sigma = \frac{P}{4}$ be the normalized ground state projector (suppose the system is on a torus with $4$ ground states). The Kraus operators of $\mathcal{E}$ are $M_1 = \sqrt{1-p}$ and $M_2 = \sqrt{p} Z$, so the action of $\mathcal{E}$ on $\sigma$ is
\beq
\mathcal{E}(\sigma) = (1-p)\frac{P}{4} + p \frac{ZPZ}{4}.
\eeq
The (generalized) inverse of this state is simple since the two terms are block diagonal and do not interfere; the result is
\beq
\mathcal{E}(\sigma)^{-1} = \frac{4}{1-p} P + \frac{4}{p} ZPZ.
\eeq

Turning to the recovery channel, the Kraus operators are
\beq
\tilde{M}_i = \sigma^{1/2} M_i^\dagger \mathcal{E}(\sigma)^{-1/2} = \frac{P}{2} M_i \left(\frac{2}{\sqrt{1-p}} P + \frac{2}{\sqrt{p}} Z P Z\right).
\eeq
This gives
\beq
\tilde{M}_1 = P
\eeq
and
\beq
\tilde{M}_2 = P Z,
\eeq
so $\mathcal{R}$ acting on $\mathcal{E}(\sigma)$ is
\beq
\mathcal{R}\left((1-p)\frac{P}{4} + p \frac{ZPZ}{4}\right) = (1-p) \frac{P}{4} + p \frac{P}{4} = \frac{P}{4}.
\eeq

But this is not true just for $P$ but for any ground state. Let $\ket{0}\bra{0}$ be a particular pure ground state. Then
\beq
\mathcal{R}(\mathcal{E}(\ket{0}\bra{0})) = \mathcal{R}( (1-p) \ket{0}\bra{0} + p Z \ket{0}\bra{0} Z) = \ket{0}\bra{0}.
\eeq
This is a toy version of the construction demonstrating the existence of an error recovery channel for the toric code.

\subsection{Review of quantum Markov chains}
\label{subsec:qmarkov}

Here we review the physics of quantum Markov chains, see \eg~\cite{qmarkov}. We say a tripartite state $\rho_{ABC}$ forms a quantum Markov chain if the conditional mutual information $I(A:C|B) = S(AB)+S(BC)-S(B)-S(ABC)$ vanishes. Classically this would imply that the probability distribution factorizes, $p(a,b,c) = \frac{p(a,b)p(b,c)}{p(b)}$, so that $A$ is independent of $C$ given $B$: $p(a,c|b) =  \frac{p(a,b,c)}{p(b)} = \frac{p(a,b)}{p(b)} \frac{p(b,c)}{p(b)}$. What is the quantum version of this statement?

Notice that strong subadditivity implies that $I(A:C|B) \geq 0$, so states with $I(A:C|B)=0$ saturate strong subadditivity. Suppose $\rho_{ABC}$ saturates strong subadditivity and consider a generic perturbation $\rho_{ABC} \rightarrow \rho_{ABC} + \delta \rho_{ABC}$. Demanding that the trace of the perturbed state be $1$ gives $\text{tr}(\delta \rho_{ABC})=0$. Now consider the variation of $I(A:C|B)$ and use $\delta S(\sigma) = - \tr( \delta \sigma \log \sigma) - \tr(\delta \sigma) = - \tr(\delta \sigma \log \sigma)$. We find
\beq
\delta I(A:C|B) = \tr\(\delta \rho_{ABC} \left[-\log \rho_{AB} - \log \rho_{BC} + \log \rho_B + \log \rho_{ABC} \right]\),
\eeq
but $I(A:C|B)$ must remain positive for all $\delta \rho$ so the linear term in the variation must vanish,
\beq
\log \rho_{ABC} = \log \rho_{AB} + \log \rho_{BC} - \log \rho_B,
\eeq
which is the quantum analog of $p(a,b,c) = \frac{p(a,b)p(b,c)}{p(b)}$.
Note that
expressions like $-\log \rho_{BC}$ should be understood as $- I_A \otimes \log \rho_{BC}$ as appropriate.
Also, beware that the various entanglement Hamiltonians $\log \rrho_{\cdot\cdot\cdot}$ here
need not commute.

What about reconstruction? In the classical case we can exactly reconstruct $p(a,b,c)$ from $p(a,b)$ and $p(b,c)$ if the conditional mutual information is zero. The reconstruction is Bayes' rule,
\beq \label{eq:bayes}
p(a,b,c) = p(c|a,b) p(a,b) \underbrace{\rightarrow}_{I=0} p(c|b) p(a,b) = \frac{p(b,c) p(a,b)}{p(b)}.
\eeq
The quantum analog of this statement is Petz's recovery channel.

Consider a quantum channel $\tr_C$ which maps $ABC$ to $AB$ by simply tracing out $C$. This is a valid quantum operation, namely throwing away a system. What is the transpose channel (see previous subsection) of the channel $\tr_C$ relative to $\sigma_{ABC} = \rho_{A} \otimes \rho_{BC}$? The Kraus operators (they are not square matrices) of $\tr_C$ are
\beq
M_c = \langle c |,
\eeq
so the Kraus operators of the Petz channel $\CR$ (which maps $AB$ to $ABC$) are
\beq
M^{\CR}_c = \rho_{A}^{1/2} \rho_{BC}^{1/2} | c \rangle \rho_{A}^{-1/2} \rho_{B}^{-1/2}.
\eeq
Note how $\rho_A$ cancels out, leaving
\beq
M^{\CR}_c = \rho_{BC}^{1/2} \rho_{B}^{-1/2} | c \rangle.
\eeq
These Kraus operators depend only on $\rho_{BC}$.

Let us compute the action of $\CR$ on a state $\sigma_{AB}$. Using the Kraus operators just derived we have
\beq
\CR(\sigma_{AB}) = \sum_c \rho_{BC}^{1/2} \rho_B^{-1/2} | c\rangle \sigma_{AB} \langle c | \rho_B^{-1/2} \rho_{BC}^{1/2}.
\eeq
The sum over $c$ can be performed immediately to yield $I_C = \sum_c \ket{c} \bra{c}$, so the channel action is
\beq
\label{eq:petz-map}
\CR(\sigma_{AB}) = \rho_{BC}^{1/2} \rho_B^{-1/2} \sigma_{AB} \rho_B^{-1/2} \rho_{BC}^{1/2}
\eeq
which is the Petz map. Note the similarity to the classical reconstruction formula \eqref{eq:bayes}. One can verify that $\CR(\rho_{AB}) = \rho_{ABC}$ using the properties of the quantum Markov chain.

We emphasize one point: every pair $\CE,\sigma$ defines a recovery channel that reverses the action of $\CE$ on $\sigma$. What is highly non-trivial here, and essential for the constructions in the main text, is that the Petz map doesn't depend on the state of the whole system but only on the state of the marginals. Combined with the vanishing of the mutual information this demonstrates that the channels above are local.

The recent developments in this field were initiated by Fawzi and Renner \cite{fr1} who showed that if $I(A:C|B) \approx 0$ then there is an approximate recovery map which acts only on $B \rightarrow BC$. Furthermore, it was later shown \cite{fr2} that this map is independent of $\rho_A$ meaning not only does it not act on the $A$ system, the $B\rightarrow BC$ map is also independent of the state on $A$. This is the machinery we use in our construction.





\section{Another construction of thermal double Hamiltonians}

This section starts from an idea of Klich and Feiguin \cite{2013arXiv1308.0756F}
which in turn was derived from the earlier literature on what we now call ``square root states".
Given a Hamiltonian $H$ and a thermal state $\rho(T)= \exp(-H/T)/Z$, \cite{2013arXiv1308.0756F}
 show that we can immediately produce a ``Hamiltonian" with the thermal double state as its exact ground state. There are two caveats: the Hamiltonian may not be local and is definitely not Hermitian.

Introduce a second copy of the system and suppose the degrees of freedom reside on sites labelled by $r$ ($r=1,...,L$). Supposing each site consists of a finite dimensional Hilbert space of dimension $\chi$ then the infinite temperature thermal double state is simply
\beq
|\infty\rangle = \left(\frac{1}{\sqrt{\chi}} \sum_i |i i \rangle \right)^L.
\eeq
This state is the unique ground state of
\beq
H_\infty = \sum_r H_{r,\infty} = \sum_r \left(1 - |\Phi\rangle\langle \Phi |_r \right)
\eeq
where
\beq
|\Phi \rangle = \frac{1}{\sqrt{\chi}} \sum_i |i i \rangle.
\eeq

The finite temperature thermal double state can be obtained from $|\infty\rangle$ by applying the square root of $\rho(T)$,
\beq
|T\rangle = (\sqrt{\rho(T)} \otimes 1) |\infty\rangle =  \left(\sqrt{\frac{e^{-\beta H}}{Z}}\otimes 1\right)|\infty\rangle.
\eeq
One easily checks that tracing out the auxiliary degrees of freedom reproduces $\rho(T)$.

$|T\rangle$ is itself the exact ground state of a set of non-Hermitian constraints given by
\beq
\tilde{H}_{r,T} = \left(e^{-\frac{H}{2T}}\otimes 1 \right) H_{r,\infty} \left( e^{\frac{H}{2T}} \otimes 1 \right).
\eeq
The Hamiltonian $\tilde{H}_T = \sum_r \tilde{H}_{r,T}$ has integer spectrum and exactly annihilates the thermal double state, but it is potentially non-local and not Hermitian. Regarding the non-locality, if correlations in the thermal state are short-ranged then we might expect $\tilde{H}_T$ to have range set by the correlation length.

A Hermitian and frustration free Hamiltonian which also has the thermal double state as its ground state is
\beq
H_T = \sum_r \tilde{H}_{r,T}^\dagger \tilde{H}_{r,T}.
\eeq
Unfortunately, the spectrum of this Hamiltonian is not under immediate control. Furthermore, the class of Hamiltonians
\beq
H_T^{(A)} = \sum_{r,r'} H_{r,T}^\dagger A_{r,r'} H_{r',T}
\eeq
with $A_{r,r'}^\dagger = A_{r',r}$ and $A_{r,r'} \geq 0$ (or some similar condition to gaurantee $H_T^{(A)} \geq 0$) all annihilate the thermal double state. The freedom to choose the set $\{A_{r,r'}\}$ could be helpful in gapping out the spectrum of $H_T^{(A)}$.

{\bf Toric code.} To illustrate one possible use of the above ansatz, consider the case of the toric code. The Hamiltonian is
\beq
H = -\sum_v A_v - \sum_p B_p,
\eeq
where $A_v$ and $B_p$ are the usual vertex and plaquette operators. The infinite temperature thermal double Hamiltonian may be taken to be
\beq
\tilde{H}_\infty = \sum_\ell 2 - X_\ell \tilde{X}_\ell - Z_\ell \tilde{Z}_\ell
\eeq
with the tilde variables referring to the purifying degrees of freedom. To obtain the non-Hermitian constraints which annihilate the thermal double state at finite $T$, we compute
\beq
\tilde{H}_{\ell,T} = \left(e^{-\frac{H}{2T}}\otimes 1 \right) H_{\ell,\infty} \left( e^{\frac{H}{2T}} \otimes 1 \right) = 2 - e^{\beta \sum_{p \ni \ell} B_p} X_\ell \tilde{X}_\ell -
e^{\beta \sum_{v \ni \ell} A_v} Z_\ell \tilde{Z}_\ell,
\eeq
where the notation in the exponentials means (finite) sums over plaquettes or vertices than contain link $\ell$.

Note that by construction the commutator between $\tilde{H}_{\ell,T}$ and $\tilde{H}_{\ell',T}$ is zero for all $\ell$ and $\ell'$. However, the commutator between $\tilde{H}^\dagger_{\ell,T}$ and $\tilde{H}_{\ell',T}$ is non-vanishing if $\ell$ and $\ell'$ are share a vertex or a plaquette. The conjugate of $\tilde{H}_{\ell,T}$ is
\beq
\tilde{H}^\dagger_{\ell,T} = 2 - X_\ell \tilde{X}_\ell e^{\beta \sum_{p \ni \ell} B_p} -
Z_\ell \tilde{Z}_\ell e^{\beta \sum_{v \ni \ell} A_v} = \tilde{H}_{\ell,-T}.
\eeq

It would be very interesting to further analyze the spectrum of the resulting hermitian Hamiltonian
$\sum_l \tilde H_{\ell,T}^\dagger \tilde H_{\ell,T}$
and show that it has an energy gap. If it does not, perhaps some modification along the lines of $H_T^{(A)}$ does?
As shown in \cite{2008arXiv0811.0033A},
the problem can be decomposed into an electric part and a magnetic part.

\section{Cellular decomposition of $\IR^4$}

\label{appendix:table}

In the following table, $d$ is the number of spatial dimensions.  $L$
is the number of unit cells (u.c.); we must keep track of this since some
of the Betti numbers decay with $L$.
$\Sigma_k$ denotes the $k$-skeleton, \ie~the result
of assembling the $0$-cells through $k$-cells.

The little cartoons depict the analogous cells in the case of $\IR^{d=3}$.
The analysis at the first two steps is given for the general case of $\IR^d$,
since the pattern is useful.
At the $2\to 3$ step we specify $d=4$.

The obstructions in the rightmost column are
given for the case of $p=1$ and $p=2$-form gauge theory \cite{Grover:2011fa}:
$$ I^{p=1} = - b_0(\partial M) |^{A+C}_{B+ ABC}
\equiv -b_0( \partial A) - b_0 (\partial C ) + b_0 (\partial B) + b_0 (\partial ABC)
$$
$$ I^{p=2} =  b_0(\partial M) |^{A+C}_{B+ ABC}  - b_1(\partial M) |^{A+C}_{B+ ABC}$$

Mysteries are in red and answer-analysis is in green.
By this we mean that we have not determined $ y_1 \equiv b_1(\partial \Sigma_2)$
a priori, but rather have used the condition (in green) of vanishing of the obstruction
$I^{p=2}_{2\to 3}$ to determine it.

\newgeometry{margin=1cm}
\begin{landscape}
\renewcommand{\arraystretch}{1.5}

\begin{table}
\begin{center}
\begin{tabular}{|p{3cm}|p{4cm}|p{4cm}|p{2cm}|p{1.4cm}|p{8cm}}
  \hline
  &$ M $ per u.c. & $\partial M$ per u.c.  & $b_0(\partial M) $ per u.c. & $b_1(\partial M) $ per u.c. & obstructions \\
\hline
$  A_{0 \to 1}$  \parfig{.04}{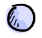}& $\Sigma_0 = $ one $B^d$ & $\partial B^d= S^{d-1}$ & 1 & 0 &
    \multirow{2}{*}{  $I_{0\to 1}^{p=1} = 2d -(d+1) = d-1, \checkmark$     }\\
\cline{1-5}
$  C_{0 \to 1}$  \parfig{.04}{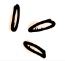}& $ d \times B^d $ & $d \times  S^{d-1}$ & $d$ & 0  &\\
\cline{1-5}
$  B_{0 \to 1}$ \parfig{.04}{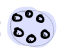}& $ 2d \times B^d $ & $2d \times  S^3$ & $2d$ & 0 &
    \multirow{2}{*}{ $I_{0\to 1}^{p=2} = - 2d + (d+1) - (d-1)  = 0 , \checkmark$ }
\\
\cline{1-5}
$  ABC_{0 \to 1}$ \parfig{.04}{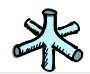}& $\Sigma_1$ & plumber's nightmare & $1/L$ & $d-1 $
&
\\
\hline
\hline
$  A_{1 \to 2}$  \parfig{.04}{figs/fig-topABC01.png} & $\Sigma_1 $ & plumber's nightmare  & $1/L$ & $d-1$ &
  \multirow{2}{*}{$I_{1\to 2}^{p=1} = - \begin{pmatrix} d \cr 2 \end{pmatrix} + \( \begin{pmatrix} d \cr 2 \end{pmatrix} + y_0 \) \sim 0 , \checkmark$  }
 \\
\cline{1-5}
$  C_{1 \to 2}$  \parfig{.04}{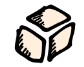}& $ \begin{pmatrix} d \cr 2 \end{pmatrix}$ 2-faces & $ \begin{pmatrix} d \cr 2 \end{pmatrix} \times  S^{d-1}$ & $ \begin{pmatrix} d \cr 2 \end{pmatrix}$& 0 &\\
\cline{1-5}
$  B_{1 \to 2}$  \parfig{.04}{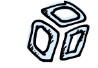} & $ \begin{pmatrix} d \cr 2 \end{pmatrix}$ donuts &
 $ \begin{pmatrix} d \cr 2 \end{pmatrix} \times S^{d-2} \times S^1$
& $ \begin{pmatrix} d \cr 2 \end{pmatrix}$ &  $ \begin{pmatrix} d \cr 2 \end{pmatrix}$  &
 \multirow{2}{*}{$I_{1\to 2}^{p=2} = - y_0 - (d-1)+ \begin{pmatrix} d \cr 2 \end{pmatrix}  + y_1
 = - 3 + 6 +
 {\cor y_1 \buildrel{?}\over{=} 7}
 \color{green} \neq 0 $ }
 \\
\cline{1-5}
$ \cor ABC_{1 \to 2}$ & $\Sigma_2$ & plumber's nightmare & $y_0 \sim 1/L$ & ${\cor y_1}$ &
\\
\hline
\hline
$  \cor A_{2 \to 3}$ & $\Sigma_2 $ & plumber's nightmare  & $1/L$ & ${\cor y_1}$  &  \multirow{2}{*}{ $I_{2\to 3}^{p=1}= - y_0  - 4 +  z_0 + 1 \sim z_0 - 3  \color{green} \neq 0 $
 }\\
\cline{1-5}
$  C_{2 \to 3}$ &  $ \begin{pmatrix} d \cr 3 \end{pmatrix}=4$ 3-volumes & $ \begin{pmatrix} d \cr 3 \end{pmatrix} \times  S^{d-1}$ & $ \begin{pmatrix} d \cr 3 \end{pmatrix}$& 0 & \\
\cline{1-5}
$ B_{2 \to 3}$ &  intersections of 3-volumes and 2-skeleton \cob  &
{  $ 4 \times S^{d-3} \times S^{2} = 4 \times S^1 \times S^2$ }
&
$z_0=4$ &
$ z_1 = 4$
&
\multirow{2}{*}{$ I_{2\to 3}^{p=2}= -I_{2\to 3}^{p=1} + y_1 - z_1
=y_1 - 4   \color{green} = 0 $  }
 \\
\cline{1-5}
$  ABC_{2 \to 3}$  \parfig{.04}{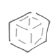}& $\Sigma_3$ = everything but the creamy filling & one $S^3$  & 1& 0 &
 \\
\hline
\hline
$  A_{3 \to 4}$  \parfig{.04}{figs/fig-topABC12.png}& $\Sigma_3 $ & one $S^3$  & 1& 0 &
\multirow{2}{*}{ $I_{3\to 4}^{p=1} = - (1+1) +2 = 0, \checkmark$ }
 \\
\cline{1-5}
$  C_{3 \to 4}$ \parfig{.04}{figs/fig-topA01.png} &  4-volume, $B^4$ & $S^3$ & 1& 0 &  \\
\cline{1-5}
$B_{3 \to 4}$ \parfig{.04}{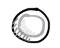}&shell around 4-volume &
$ 2 \times S^3$
& 2 & 0  &
\multirow{2}{*}{  $I_{3\to 4}^{p=2} =  0, \checkmark$ }\\
\cline{1-5}
$  ABC_{3 \to 4}$ & everything  & Nothing  & 0& 0 & \\
\hline
\end{tabular}
\end{center}
\end{table}

\renewcommand{\arraystretch}{1}

\end{landscape}

\restoregeometry

\phantomsection\addcontentsline{toc}{section}{References}
\bibliography{collection}

\begingroup\raggedright\begin{thebibliography}{10}

\bibitem{sourcery1}
B.~{Swingle} and J.~{McGreevy}, ``{Renormalization group constructions of
  topological quantum liquids and beyond},'' {\em ArXiv e-prints} (July, 2014)
  \href{http://xxx.lanl.gov/abs/1407.8203}{{\tt 1407.8203}}.

\bibitem{mera}
G.~Vidal, ``Class of Quantum Many-Body States That Can Be Efficiently
  Simulated,'' {\em Phys. Rev. Lett.} {\bf 101} (Sep, 2008) 110501,
  \href{http://link.aps.org/doi/10.1103/PhysRevLett.101.110501}{http://link.aps.org/doi/10.1103/PhysRevLett.101.110501}.

\bibitem{RevModPhys.82.1743}
B.~Spivak, S.~V. Kravchenko, S.~A. Kivelson, and X.~P.~A. Gao,
  ``\textit{Colloquium} : Transport in strongly correlated two dimensional
  electron fluids,'' {\em Rev. Mod. Phys.} {\bf 82} (May, 2010) 1743--1766,
  \href{http://link.aps.org/doi/10.1103/RevModPhys.82.1743}{http://link.aps.org/doi/10.1103/RevModPhys.82.1743}.

\bibitem{Prosen:2015xha}
T.~Prosen, ``{Matrix product solutions of boundary driven quantum chains},''
  {\em J. Phys.} {\bf A48} (2015), no.~37 373001,
  \href{http://xxx.lanl.gov/abs/1504.00783}{{\tt 1504.00783}}.

\bibitem{transport-to-appear}
B.~{Swingle}, S.~{Mumford}, R.~{Mahajan}, D.~{Freeman}, and N.~{Tubman},
  ``Entanglement structure of non-equilibrium steady states,'' {\em to appear}
  (2016).

\bibitem{2014arXiv1409.3435K}
M.~J. {Kastoryano} and F.~G.~S.~L. {Brand\~ao}, ``{Quantum Gibbs Samplers: the
  commuting case},'' {\em ArXiv e-prints} (Sept., 2014)
  \href{http://xxx.lanl.gov/abs/1409.3435}{{\tt 1409.3435}}.

\bibitem{2015PhRvB..91d5138M}
A.~{Molnar}, N.~{Schuch}, F.~{Verstraete}, and J.~I. {Cirac}, ``{Approximating
  Gibbs states of local Hamiltonians efficiently with projected entangled pair
  states},'' {\em \prb} {\bf 91} (Jan., 2015) 045138,
  \href{http://xxx.lanl.gov/abs/1406.2973}{{\tt 1406.2973}}.

\bibitem{Stanford:2014jda}
D.~Stanford and L.~Susskind, ``{Complexity and Shock Wave Geometries},'' {\em
  Phys. Rev.} {\bf D90} (2014), no.~12 126007,
  \href{http://xxx.lanl.gov/abs/1406.2678}{{\tt 1406.2678}}.

\bibitem{Brown:2015lvg}
A.~R. Brown, D.~A. Roberts, L.~Susskind, B.~Swingle, and Y.~Zhao,
  ``{Complexity, action, and black holes},'' {\em Phys. Rev.} {\bf D93} (2016),
  no.~8 086006, \href{http://xxx.lanl.gov/abs/1512.04993}{{\tt 1512.04993}}.

\bibitem{Brown:2015bva}
A.~R. Brown, D.~A. Roberts, L.~Susskind, B.~Swingle, and Y.~Zhao,
  ``{Holographic Complexity Equals Bulk Action?},'' {\em Phys. Rev. Lett.} {\bf
  116} (2016), no.~19 191301, \href{http://xxx.lanl.gov/abs/1509.07876}{{\tt
  1509.07876}}.

\bibitem{Maldacena:2001kr}
J.~M. Maldacena, ``{Eternal black holes in anti-de Sitter},'' {\em JHEP} {\bf
  04} (2003) 021, \href{http://xxx.lanl.gov/abs/hep-th/0106112}{{\tt
  hep-th/0106112}}.

\bibitem{2013arXiv1308.0756F}
A.~E. {Feiguin} and I.~{Klich}, ``{Hermitian and non-Hermitian thermal
  Hamiltonians},'' {\em ArXiv e-prints} (Aug., 2013)
  \href{http://xxx.lanl.gov/abs/1308.0756}{{\tt 1308.0756}}.

\bibitem{2012PhRvB..86x5116K}
I.~H. {Kim}, ``{Perturbative analysis of topological entanglement entropy from
  conditional independence},'' {\em \prb} {\bf 86} (Dec., 2012) 245116,
  \href{http://xxx.lanl.gov/abs/1210.2360}{{\tt 1210.2360}}.

\bibitem{2013PhRvB..87o5120K}
I.~H. {Kim}, ``{Determining the structure of the real-space entanglement
  spectrum from approximate conditional independence},'' {\em \prb} {\bf 87}
  (Apr., 2013) 155120, \href{http://xxx.lanl.gov/abs/1210.1831}{{\tt
  1210.1831}}.

\bibitem{fr1}
O.~{Fawzi} and R.~{Renner}, ``{Quantum conditional mutual information and
  approximate Markov chains},'' {\em ArXiv e-prints} (Oct., 2014)
  \href{http://xxx.lanl.gov/abs/1410.0664}{{\tt 1410.0664}}.

\bibitem{fr2}
D.~{Sutter}, O.~{Fawzi}, and R.~{Renner}, ``{Universal recovery map for
  approximate Markov chains},'' {\em ArXiv e-prints} (Apr., 2015)
  \href{http://xxx.lanl.gov/abs/1504.07251}{{\tt 1504.07251}}.

\bibitem{qmarkov}
P.~{Hayden}, R.~{Jozsa}, D.~{Petz}, and A.~{Winter}, ``{Structure of States
  Which Satisfy Strong Subadditivity of Quantum Entropy with Equality},'' {\em
  Communications in Mathematical Physics} {\bf 246} (2004) 359--374,
  \href{http://xxx.lanl.gov/abs/quant-ph/0304007}{{\tt quant-ph/0304007}}.

\bibitem{2006PhRvL..96r1602R}
S.~{Ryu} and T.~{Takayanagi}, ``{Holographic Derivation of Entanglement Entropy
  from the anti de Sitter Space/Conformal Field Theory Correspondence},'' {\em
  Physical Review Letters} {\bf 96} (May, 2006) 181602,
  \href{http://xxx.lanl.gov/abs/hep-th/0603001}{{\tt hep-th/0603001}}.

\bibitem{Swingle:2009bg}
B.~Swingle, ``{Entanglement Renormalization and Holography},'' {\em Phys.Rev.}
  {\bf D86} (2012) 065007, \href{http://xxx.lanl.gov/abs/0905.1317}{{\tt
  0905.1317}}.

\bibitem{Swingle:2012wq}
B.~Swingle, ``{Constructing holographic spacetimes using entanglement
  renormalization},'' \href{http://xxx.lanl.gov/abs/1209.3304}{{\tt
  1209.3304}}.

\bibitem{2014arXiv1412.0732E}
G.~{Evenbly} and G.~{Vidal}, ``{Tensor Network Renormalization},'' {\em Phys.
  Rev. Lett.} {\bf 115} (Oct, 2015) 180405,
  \href{http://xxx.lanl.gov/abs/1412.0732}{{\tt 1412.0732}},
  \href{http://link.aps.org/doi/10.1103/PhysRevLett.115.180405}{http://link.aps.org/doi/10.1103/PhysRevLett.115.180405}.

\bibitem{2011PhRvL.107u0501H}
M.~B. {Hastings}, ``{Topological Order at Nonzero Temperature},'' {\em Physical
  Review Letters} {\bf 107} (Nov., 2011) 210501,
  \href{http://xxx.lanl.gov/abs/1106.6026}{{\tt 1106.6026}}.

\bibitem{Kitaev-2013}
A.~Kitaev, ``On the Classification of Short-Range Entangled States,'' {\em
  unpublished} (2013)
  \href{http://scgp.stonybrook.edu/archives/7874}{http://scgp.stonybrook.edu/archives/7874}.

\bibitem{2013PhRvL.111q0501V}
K.~{Van Acoleyen}, M.~{Mari{\"e}n}, and F.~{Verstraete}, ``{Entanglement Rates
  and Area Laws},'' {\em Physical Review Letters} {\bf 111} (Oct., 2013)
  170501, \href{http://xxx.lanl.gov/abs/1304.5931}{{\tt 1304.5931}}.

\bibitem{2005PhRvB..72d5141H}
M.~B. {Hastings} and X.-G. {Wen}, ``{Quasiadiabatic continuation of quantum
  states: The stability of topological ground-state degeneracy and emergent
  gauge invariance},'' {\em \prb} {\bf 72} (July, 2005) 045141,
  \href{http://xxx.lanl.gov/abs/cond-mat/0503554}{{\tt cond-mat/0503554}}.

\bibitem{2004PhRvB..69j4431H}
M.~B. {Hastings}, ``{Lieb-Schultz-Mattis in higher dimensions},'' {\em \prb}
  {\bf 69} (Mar., 2004) 104431,
  \href{http://xxx.lanl.gov/abs/cond-mat/0305505}{{\tt cond-mat/0305505}}.

\bibitem{2010arXiv1008.5137H}
M.~B. {Hastings}, ``{Locality in Quantum Systems},'' {\em ArXiv e-prints}
  (Aug., 2010) \href{http://xxx.lanl.gov/abs/1008.5137}{{\tt 1008.5137}}.

\bibitem{2013Swingle-Senthil}
B.~{Swingle} and T.~{Senthil}, ``{Universal crossovers between entanglement
  entropy and thermal entropy},'' {\em \prb} {\bf 87} (Jan., 2013) 045123,
  \href{http://xxx.lanl.gov/abs/1112.1069}{{\tt 1112.1069}}.

\bibitem{Calabrese:2009qy}
P.~Calabrese and J.~Cardy, ``{Entanglement entropy and conformal field
  theory},'' {\em J.Phys.} {\bf A42} (2009) 504005,
  \href{http://xxx.lanl.gov/abs/0905.4013}{{\tt 0905.4013}}.

\bibitem{prettygoodchannel}
H.~{Barnum} and E.~{Knill}, ``{Reversing quantum dynamics with near-optimal
  quantum and classical fidelity},'' {\em eprint arXiv:quant-ph/0004088} (Apr.,
  2000) \href{http://xxx.lanl.gov/abs/quant-ph/0004088}{{\tt
  quant-ph/0004088}}.

\bibitem{Swingle:2010jz}
B.~Swingle, ``{Mutual information and the structure of entanglement in quantum
  field theory},'' \href{http://xxx.lanl.gov/abs/1010.4038}{{\tt 1010.4038}}.

\bibitem{Hung:2014npa}
L.-Y. Hung, R.~C. Myers, and M.~Smolkin, ``{Twist operators in higher
  dimensions},'' {\em JHEP} {\bf 10} (2014) 178,
  \href{http://xxx.lanl.gov/abs/1407.6429}{{\tt 1407.6429}}.

\bibitem{Hung:2011nu}
L.-Y. Hung, R.~C. Myers, M.~Smolkin, and A.~Yale, ``{Holographic Calculations
  of Renyi Entropy},'' {\em JHEP} {\bf 12} (2011) 047,
  \href{http://xxx.lanl.gov/abs/1110.1084}{{\tt 1110.1084}}.

\bibitem{Bueno:2015qya}
P.~Bueno, R.~C. Myers, and W.~Witczak-Krempa, ``{Universal corner entanglement
  from twist operators},'' \href{http://xxx.lanl.gov/abs/1507.06997}{{\tt
  1507.06997}}.

\bibitem{Dennis:2001nw}
E.~Dennis, A.~Kitaev, A.~Landahl, and J.~Preskill, ``{Topological quantum
  memory},'' {\em J. Math. Phys.} {\bf 43} (2002) 4452--4505,
  \href{http://xxx.lanl.gov/abs/quant-ph/0110143}{{\tt quant-ph/0110143}}.

\bibitem{Bueno:2015rda}
P.~Bueno, R.~C. Myers, and W.~Witczak-Krempa, ``{Universality of corner
  entanglement in conformal field theories},'' {\em Phys. Rev. Lett.} {\bf 115}
  (2015), no.~2 021602, \href{http://xxx.lanl.gov/abs/1505.04804}{{\tt
  1505.04804}}.

\bibitem{Faulkner:2013ana}
T.~Faulkner, A.~Lewkowycz, and J.~Maldacena, ``{Quantum corrections to
  holographic entanglement entropy},'' {\em JHEP} {\bf 11} (2013) 074,
  \href{http://xxx.lanl.gov/abs/1307.2892}{{\tt 1307.2892}}.

\bibitem{Hubeny-Swingle}
V.~Hubeny and B.~Swingle, ``work in progress,''.

\bibitem{1999IJTP...38.1113M}
J.~{Maldacena}, ``{The Large-N Limit of Superconformal Field Theories and
  Supergravity},'' {\em International Journal of Theoretical Physics} {\bf 38}
  (1999) 1113--1133, \href{http://xxx.lanl.gov/abs/hep-th/9711200}{{\tt
  hep-th/9711200}}.

\bibitem{1998PhLB..428..105G}
S.~S. {Gubser}, I.~R. {Klebanov}, and A.~M. {Polyakov}, ``{Gauge theory
  correlators from non-critical string theory},'' {\em Physics Letters B} {\bf
  428} (May, 1998) 105--114, \href{http://xxx.lanl.gov/abs/hep-th/9802109}{{\tt
  hep-th/9802109}}.

\bibitem{1998AdTMP...2..253W}
E.~{Witten}, ``{Anti-de Sitter space and holography},'' {\em Advances in
  Theoretical and Mathematical Physics} {\bf 2} (1998) 253,
  \href{http://xxx.lanl.gov/abs/hep-th/9802150}{{\tt hep-th/9802150}}.

\bibitem{Aharony:1999ti}
O.~Aharony, S.~S. Gubser, J.~M. Maldacena, H.~Ooguri, and Y.~Oz, ``{Large N
  field theories, string theory and gravity},'' {\em Phys. Rept.} {\bf 323}
  (2000) 183--386, \href{http://xxx.lanl.gov/abs/hep-th/9905111}{{\tt
  hep-th/9905111}}.

\bibitem{Maldacena:2003nj}
J.~M. Maldacena, ``{TASI 2003 lectures on AdS / CFT},'' in {\em {Progress in
  string theory. Proceedings, Summer School, TASI 2003, Boulder, USA, June
  2-27, 2003}}, pp.~155--203, 2003.
\newblock \href{http://xxx.lanl.gov/abs/hep-th/0309246}{{\tt hep-th/0309246}}.

\bibitem{Horowitz:2006ct}
G.~T. Horowitz and J.~Polchinski, ``{Gauge/gravity duality},''
  \href{http://xxx.lanl.gov/abs/gr-qc/0602037}{{\tt gr-qc/0602037}}.

\bibitem{Hartnoll:2009sz}
S.~A. Hartnoll, ``{Lectures on holographic methods for condensed matter
  physics},'' {\em Class. Quant. Grav.} {\bf 26} (2009) 224002,
  \href{http://xxx.lanl.gov/abs/0903.3246}{{\tt 0903.3246}}.

\bibitem{2009arXiv0909.3553H}
S.~A. {Hartnoll}, ``{Quantum Critical Dynamics from Black Holes},''
  \href{http://xxx.lanl.gov/abs/0909.3553}{{\tt 0909.3553}}.

\bibitem{McGreevy:2009xe}
J.~McGreevy, ``{Holographic duality with a view toward many-body physics},''
  {\em Adv.High Energy Phys.} {\bf 2010} (2010) 723105,
  \href{http://xxx.lanl.gov/abs/0909.0518}{{\tt 0909.0518}}.

\bibitem{Swingle:2015ipa}
B.~Swingle and J.~McGreevy, ``{Area Law for Gapless States from Local
  Entanglement Thermodynamics},'' {\em Phys. Rev. B} {\bf 93} (May, 2015)
  205120, \href{http://xxx.lanl.gov/abs/1505.07106}{{\tt 1505.07106}}.

\bibitem{PhysRevB.81.235102}
G.~Evenbly and G.~Vidal, ``Entanglement renormalization in noninteracting
  fermionic systems,'' {\em Phys. Rev. B} {\bf 81} (Jun, 2010) 235102,
  \href{http://link.aps.org/doi/10.1103/PhysRevB.81.235102}{http://link.aps.org/doi/10.1103/PhysRevB.81.235102}.

\bibitem{2015WhiteFreeFermions}
M.~T. {Fishman} and S.~R. {White}, ``{Compression of correlation matrices and
  an efficient method for forming matrix product states of fermionic Gaussian
  states},'' {\em \prb} {\bf 92} (Aug., 2015) 075132,
  \href{http://xxx.lanl.gov/abs/1504.07701}{{\tt 1504.07701}}.

\bibitem{Evenbly:2016cly}
G.~Evenbly and S.~R. White, ``{Entanglement renormalization and wavelets},''
  {\em Phys. Rev. Lett.} {\bf 116} (2016), no.~14 140403,
  \href{http://xxx.lanl.gov/abs/1602.01166}{{\tt 1602.01166}}.

\bibitem{Czech:2015xna}
B.~Czech, G.~Evenbly, L.~Lamprou, S.~McCandlish, X.-L. Qi, J.~Sully, and
  G.~Vidal, ``{A tensor network quotient takes the vacuum to the thermal
  state},'' \href{http://xxx.lanl.gov/abs/1510.07637}{{\tt 1510.07637}}.

\bibitem{sqrt-draft}
B.~{Swingle}, J.~{McGreevy}, and S.~{Xu}, ``{Renormalization group circuits for
  gapless states},'' {\em Phys. Rev. B} {\bf 93} (Feb., 2016) 205159,
  \href{http://xxx.lanl.gov/abs/1602.02805}{{\tt 1602.02805}},
  \href{http://link.aps.org/doi/10.1103/PhysRevB.93.205159}{http://link.aps.org/doi/10.1103/PhysRevB.93.205159}.

\bibitem{2007PhRvB..76r4442C}
C.~{Castelnovo} and C.~{Chamon}, ``{Entanglement and topological entropy of the
  toric code at finite temperature},'' {\em \prb} {\bf 76} (Nov., 2007) 184442,
  \href{http://xxx.lanl.gov/abs/0704.3616}{{\tt 0704.3616}}.

\bibitem{2008arXiv0811.0033A}
R.~{Alicki}, M.~{Horodecki}, P.~{Horodecki}, and R.~{Horodecki}, ``{On thermal
  stability of topological qubit in Kitaev's 4D model},'' {\em ArXiv e-prints}
  (Nov., 2008) \href{http://xxx.lanl.gov/abs/0811.0033}{{\tt 0811.0033}}.

\bibitem{Mazac:2011np}
D.~Mazac and A.~Hamma, ``{Topological order, entanglement, and quantum memory
  at finite temperature},'' {\em Annals Phys.} {\bf 327} (2012) 2096,
  \href{http://xxx.lanl.gov/abs/1112.0947}{{\tt 1112.0947}}.

\bibitem{Grover:2011fa}
T.~Grover, A.~M. Turner, and A.~Vishwanath, ``{Entanglement Entropy of Gapped
  Phases and Topological Order in Three dimensions},'' {\em Phys.Rev.} {\bf
  B84} (2011) 195120, \href{http://xxx.lanl.gov/abs/1108.4038}{{\tt
  1108.4038}}.

\bibitem{2014PhRvL.112g0501H}
M.~B. {Hastings}, G.~H. {Watson}, and R.~G. {Melko}, ``{Self-Correcting Quantum
  Memories Beyond the Percolation Threshold},'' {\em Physical Review Letters}
  {\bf 112} (Feb., 2014) 070501, \href{http://xxx.lanl.gov/abs/1309.2680}{{\tt
  1309.2680}}.

\bibitem{Atick:1988si}
J.~J. Atick and E.~Witten, ``{The Hagedorn Transition and the Number of Degrees
  of Freedom of String Theory},'' {\em Nucl. Phys.} {\bf B310} (1988) 291--334.

\bibitem{Horowitz:1996nw}
G.~T. Horowitz and J.~Polchinski, ``{A Correspondence principle for black holes
  and strings},'' {\em Phys. Rev.} {\bf D55} (1997) 6189--6197,
  \href{http://xxx.lanl.gov/abs/hep-th/9612146}{{\tt hep-th/9612146}}.

\bibitem{Horowitz:1997jc}
G.~T. Horowitz and J.~Polchinski, ``{Selfgravitating fundamental strings},''
  {\em Phys. Rev.} {\bf D57} (1998) 2557--2563,
  \href{http://xxx.lanl.gov/abs/hep-th/9707170}{{\tt hep-th/9707170}}.

\bibitem{2015RSPSA.47150338W}
M.~M. {Wilde}, ``{Recoverability in quantum information theory},'' {\em
  Proceedings of the Royal Society of London Series A} {\bf 471} (Oct., 2015)
  20150338, \href{http://xxx.lanl.gov/abs/1505.04661}{{\tt 1505.04661}}.

\bibitem{2015arXiv150907127J}
M.~{Junge}, R.~{Renner}, D.~{Sutter}, M.~M. {Wilde}, and A.~{Winter},
  ``{Universal recovery from a decrease of quantum relative entropy},'' {\em
  ArXiv e-prints} (Sept., 2015) \href{http://xxx.lanl.gov/abs/1509.07127}{{\tt
  1509.07127}}.

\bibitem{2012LNP...843..245A}
R.~{Augusiak}, F.~M. {Cucchietti}, and M.~{Lewenstein}, ``{Many-Body Physics
  from a Quantum Information Perspective},'' in {\em Lecture Notes in Physics,
  Berlin Springer Verlag} (D.~C. {Cabra}, A.~{Honecker}, and P.~{Pujol}, eds.),
  vol.~843 of {\em Lecture Notes in Physics, Berlin Springer Verlag}, p.~245,
  2012.
\newblock \href{http://xxx.lanl.gov/abs/1003.3153}{{\tt 1003.3153}}.

\bibitem{2006Sci...311.1133N}
M.~A. {Nielsen}, M.~R. {Dowling}, M.~{Gu}, and A.~C. {Doherty}, ``{Quantum
  Computation as Geometry},'' {\em Science} {\bf 311} (Feb., 2006) 1133--1135,
  \href{http://xxx.lanl.gov/abs/quant-ph/0603161}{{\tt quant-ph/0603161}}.

\bibitem{2011PhRvA..83d2330H}
J.~{Haah}, ``{Local stabilizer codes in three dimensions without string logical
  operators},'' {\em \pra} {\bf 83} (Apr., 2011) 042330,
  \href{http://xxx.lanl.gov/abs/1101.1962}{{\tt 1101.1962}}.

\bibitem{2014PhRvB..89g5119H}
J.~{Haah}, ``{Bifurcation in entanglement renormalization group flow of a
  gapped spin model},'' {\em \prb} {\bf 89} (Feb., 2014) 075119,
  \href{http://xxx.lanl.gov/abs/1310.4507}{{\tt 1310.4507}}.

\bibitem{PhysRevLett.107.150504}
S.~Bravyi and J.~Haah, ``Energy Landscape of 3D Spin Hamiltonians with
  Topological Order,'' {\em Phys. Rev. Lett.} {\bf 107} (Oct, 2011) 150504,
  \href{http://link.aps.org/doi/10.1103/PhysRevLett.107.150504}{http://link.aps.org/doi/10.1103/PhysRevLett.107.150504}.

\bibitem{2013PhDHaah}
J.~{Haah}, {\em {Lattice quantum codes and exotic topological phases of
  matter}}.
\newblock PhD thesis, California Institute of Technology, 2013.
\newblock \href{http://xxx.lanl.gov/abs/1305.6973}{{\tt 1305.6973}}.

\bibitem{PhysRevLett.111.200501}
S.~Bravyi and J.~Haah, ``Quantum Self-Correction in the 3D Cubic Code Model,''
  {\em Phys. Rev. Lett.} {\bf 111} (Nov, 2013) 200501,
  \href{http://link.aps.org/doi/10.1103/PhysRevLett.111.200501}{http://link.aps.org/doi/10.1103/PhysRevLett.111.200501}.

\bibitem{2016arXiv160307805S}
K.~{Siva} and B.~{Yoshida}, ``{Topological Order and Memory Time in Marginally
  Self-Correcting Quantum Memory},'' {\em ArXiv e-prints} (Mar., 2016)
  \href{http://xxx.lanl.gov/abs/1603.07805}{{\tt 1603.07805}}.

\bibitem{2016arXiv160405329P}
M.~{Pretko}, ``{Subdimensional Particle Structure of Higher Rank U(1) Spin
  Liquids},'' {\em ArXiv e-prints} (Apr., 2016)
  \href{http://xxx.lanl.gov/abs/1604.05329}{{\tt 1604.05329}}.

\bibitem{2016arXiv160608857P}
M.~{Pretko}, ``{Generalized Electromagnetism of Subdimensional Particles: A
  Spin Liquid Story},'' {\em ArXiv e-prints} (June, 2016)
  \href{http://xxx.lanl.gov/abs/1606.08857}{{\tt 1606.08857}}.

\bibitem{2012PhRvL.109t7202B}
S.~{Bravyi}, L.~{Caha}, R.~{Movassagh}, D.~{Nagaj}, and P.~W. {Shor},
  ``{Criticality without Frustration for Quantum Spin-1 Chains},'' {\em
  Physical Review Letters} {\bf 109} (Nov., 2012) 207202,
  \href{http://xxx.lanl.gov/abs/1203.5801}{{\tt 1203.5801}}.

\bibitem{2014arXiv1408.1657M}
R.~{Movassagh} and P.~W. {Shor}, ``{Power law violation of the area law in
  critical spin chains},'' {\em ArXiv e-prints} (Aug., 2014)
  \href{http://xxx.lanl.gov/abs/1408.1657}{{\tt 1408.1657}}.

\bibitem{2016arXiv160503842S}
O.~{Salberger} and V.~{Korepin}, ``{Fredkin Spin Chain},'' {\em ArXiv e-prints}
  (May, 2016) \href{http://xxx.lanl.gov/abs/1605.03842}{{\tt 1605.03842}}.

\bibitem{petz1986}
D.~Petz, ``Sufficient subalgebras and the relative entropy of states of a von
  Neumann algebra,'' {\em Comm. Math. Phys.} {\bf 105} (1986), no.~1 123--131,
  \href{http://projecteuclid.org/euclid.cmp/1104115260}{http://projecteuclid.org/euclid.cmp/1104115260}.

\bibitem{PETZ01031988}
D.~PETZ, ``SUFFICIENCY OF CHANNELS OVER VON NEUMANN ALGEBRAS,'' {\em The
  Quarterly Journal of Mathematics} {\bf 39} (1988), no.~1 97--108,
  \href{http://xxx.lanl.gov/abs/http://qjmath.oxfordjournals.org/content/39/1/97.full.pdf+html}{{\tt
  http://qjmath.oxfordjournals.org/content/39/1/97.full.pdf+html}},
  \href{http://qjmath.oxfordjournals.org/content/39/1/97.short}{http://qjmath.oxfordjournals.org/content/39/1/97.short}.

\bibitem{2003RvMaP..15...79P}
D.~{Petz}, ``{Monotonicity of Quantum Relative Entropy Revisited},'' {\em
  Reviews in Mathematical Physics} {\bf 15} (2003) 79--91,
  \href{http://xxx.lanl.gov/abs/quant-ph/0209053}{{\tt quant-ph/0209053}}.

\bibitem{2003AnPhy.303....2K}
A.~Y. {Kitaev}, ``{Fault-tolerant quantum computation by anyons},'' {\em Annals
  of Physics} {\bf 303} (Jan., 2003) 2--30,
  \href{http://xxx.lanl.gov/abs/quant-ph/9707021}{{\tt quant-ph/9707021}}.

\end{thebibliography}\endgroup
\bibliographystyle{ucsd}

\end{document}